\documentclass[a4paper,fleqn]{cas-sc}

\usepackage{xcolor}
\usepackage{makecell}
\definecolor{darkgreen}{rgb}{0.0, 0.86, 0.0} % Darker green
\DeclareRobustCommand{\rev}[1]{#1}
\DeclareRobustCommand{\revred}[1]{#1}
\DeclareRobustCommand{\revgreen}[1]{#1}
\DeclareRobustCommand{\revorange}[1]{#1}

\usepackage{amsmath}
\usepackage{amssymb}
\usepackage[numbers]{natbib}
\usepackage{subcaption}
\usepackage{caption}
\usepackage{multirow}
\graphicspath{{Figs/}{Final_figures/}{Final_figures/Experiments/}{Final_figures/Efficiency/}}
\usepackage[switch]{lineno}
\def\tsc#1{\csdef{#1}{\textsc{\lowercase{#1}}\xspace}}
\tsc{WGM}
\tsc{QE}
\tsc{EP}
\tsc{PMS}
\tsc{BEC}
\tsc{DE}
\begin{document}
\ExplSyntaxOn
\cs_gset:Npn \__first_footerline: {}
\ExplSyntaxOff
\let\WriteBookmarks\relax
\def\floatpagepagefraction{1}
\def\textpagefraction{.001}

% Short title
\shorttitle{KATOsuper: Surrogate-accelerated neural topology optimization}

% Short author
\shortauthors{Shengyu Yan et~al.}

% Main title of the paper
\title [mode = title]{KATOsuper: Surrogate-accelerated neural topology optimization with sensitivity-consistent Fourier neural operators}

% First author
\author[1]{Shengyu Yan}[orcid=0000-0002-9077-9649]

% Corresponding author
\author[1,2]{Jasmin Jelovica}[orcid=0000-0002-8396-941X]
\cormark[1]
\ead{jjelovica@mech.ubc.ca}

\affiliation[1]{organization={Department of Mechanical Engineering, The University of British Columbia},
    addressline={6250 Applied Science Ln}, 
    city={Vancouver},
    postcode={V6T 1Z4}, 
    country={Canada}}

\affiliation[2]{organization={Department of Civil Engineering, The University of British Columbia},
    addressline={6250 Applied Science Ln}, 
    city={Vancouver},
    postcode={V6T 1Z4}, 
    country={Canada}}

\cortext[cor1]{Corresponding author}

\begin{abstract}
Topology optimization (TO) remains computationally intensive due to repeated finite element analysis (FEA) evaluations required at each iteration. While neural network-based surrogates offer potential acceleration, existing approaches often suffer from gradient inconsistency between predicted objectives and sensitivities, leading to optimization instability. This work presents \textbf{KATOsuper}, an objective-agnostic framework that couples neural-reparameterized topology optimization with a Sensitivity-Consistent Fourier Neural Operator (SC-FNO). The framework employs the \texttt{forward\_split} architecture, which derives deployed sensitivities via automatic differentiation through the predicted objective field and thereby preserves consistency between the predicted objective and the gradient used for optimization. \revorange{The case studies include three 2D benchmark problems and three 3D structures considering compliance or stress minimization.} A physics-informed multi-channel input encoding with Fourier position embedding enables resolution-invariant learning, supporting zero-shot extrapolation beyond the training resolution, with useful performance at moderate scaling factors and topology-preserving exploration at up to 64$\times$ without retraining. The framework extends to 3D through KATO3D, featuring novel KANConv3D blocks with learnable B-spline activations. KATOsuper demonstrates \revorange{15--110$\times$ deployment-time speedup} over MATLAB baselines while \revgreen{maintaining competitive optimality, with the clearest gains observed in complex 3D and stress-optimization cases}. The insight that sensitivity \textit{direction} matters more than magnitude enables robust optimization even with approximate physics evaluation---extensible to other differentiable physics-driven design objectives.
\end{abstract}

%\begin{graphicalabstract}
%    \begin{center}
%    \includegraphics[width=\textwidth]%{Final_figures/GA.png}
%    \end{center}
%\end{graphicalabstract}

% Keywords

\maketitle

\section{Introduction}\label{intro}

Topology optimization (TO) seeks the optimal material distribution within a design domain to minimize an objective function subject to constraints. Since the foundational work of Bendsøe and Kikuchi \cite{bendsoe1988generating,bendsoe1989optimal} on homogenization-based TO, the field has evolved through density-based methods \cite{bendsoe2003topology,sigmund2013topology}, level-set formulations \cite{wang2003level,allaire2004structural}, and evolutionary approaches \cite{xie1993simple,huang2010evolutionary}, as comprehensively reviewed by Shin et al. \cite{shin2023topology}. The Solid Isotropic Material with Penalization (SIMP) method \cite{bendsoe2003topology} is widely adopted, with Sigmund's 99-line MATLAB code \cite{sigmund200199} and its 3D extension \cite{liu2014efficient} facilitating broad use across academia and industry. Classical optimizers including Optimality Criteria (OC) \cite{zhou1991coc} and the Method of Moving Asymptotes (MMA) \cite{svanberg1987method} provide efficient gradient-based updates, while filtering techniques \cite{sigmund1998numerical} suppress checkerboard patterns.

Despite decades of progress, large-scale topology optimization faces fundamental challenges that limit industrial adoption, as reviewed by Mukherjee et al. \cite{mukherjee_accelerating_2021}. The primary challenge is the prohibitive computational cost of solving finite element equations at each iteration. For fine 3D meshes, a single FEA solve can take seconds to minutes, and complete optimization requires hundreds of such solves \cite{yago2022topology}. Giga-voxel scale optimization has been demonstrated \cite{aage2017giga}, but requires supercomputing resources beyond typical industrial access. The second challenge is the tight coupling between analysis and optimization, making it impossible to pre-compute or cache expensive computations. For stress-based objectives, the computational burden intensifies due to non-smooth p-norm aggregation \cite{le2010stress}, ill-conditioned sensitivities \cite{holmberg2013stress}, and singular optima requiring careful relaxation \cite{cheng1997varepsilon,bruggi2008mixed,deng2021efficient}. A typical 2D optimization at a resolution of $128 \times 128$ requires hundreds of finite element solves,
whereas moderate-resolution 3D problems may require hours or even days of computation. Real-world applications demand much higher element resolutions, which further increases the computational cost.

\rev{Significant efforts have been devoted to accelerating FEA within TO loops through classical approaches including re-analysis, multi-grid solvers, and model reduction \cite{mukherjee_accelerating_2021}. High-performance computing frameworks have emerged leveraging parallel architectures: PETSc-based topology optimization \cite{aage2015topology} provides fully parallel computing on distributed clusters, JAX-FEM \cite{xue2023jax} offers differentiable GPU-accelerated 3D FEA with automatic differentiation, and FEniTop \cite{jia2024fenitop} supports 2D and 3D parallel computing via FEniCSx. While these frameworks achieve impressive performance, they are typically associated with Linux-based systems requiring MPI, specialized compilers, and complex installation procedures, which may limit accessibility for users outside Linux-based HPC environments.}

Machine learning offers data-driven approaches to bypass expensive physics simulations. Early neural network applications focused on surrogate modeling and meta-learning \cite{yildiz2003integrated,chi2021universal} or generative design synthesis \cite{oh2019deep,li2019non}. Convolutional neural networks (CNNs) have been applied to super-resolution \cite{wang2021deep} and high-resolution topology prediction \cite{xue2021efficient}. A significant development was the introduction of neural reparameterization, inspired by the Deep Image Prior (DIP) \cite{ulyanov2018deep}. DIP demonstrates that the architecture of CNNs itself imposes a strong implicit prior toward natural image statistics, even without any training data. When an untrained CNN is optimized to fit a target through gradient descent, the network tends to generate structured patterns. \revgreen{Recent ML-assisted topology optimization has also used learned components inside optimizers, including CNN-accelerated numerical-shape-function prediction with uncertainty clustering for robust level-set/multiscale FEM topology optimization~\cite{li2026cmame_multiscale_uncertainty}, and RBF-NN/particle-swarm load-location optimization in robust concurrent topology--device-layout design~\cite{li2026ast_device_layout}.} Hoyer et al. \cite{hoyer2019neural} first applied this to TO, using untrained CNNs to parameterize density fields. Subsequent work extended this to multi-physics problems \cite{zhang2021tonr,chandrasekhar2021tounn} with open-source implementations \cite{erzmann2023dl4to,liu2026topology}. Our prior work KATO \cite{yan2025kato} introduced convolutional KANs \cite{bodner2024convolutional,liu2024kan} for stress minimization by enhancing the implicit prior capability introduced by DIP \cite{ulyanov2018deep}. However, these methods still require FEA at every step, leaving the primary computational bottleneck unaddressed.

\rev{Surrogate models offer a path to bypass FEA by predicting structural responses directly from material distributions. Physics-informed neural networks \cite{raissi2019physics,haghighat2021physics,jeong_physics-informed_2023} embed physical equations but struggle with high-contrast materials typical in TO. Neural operators---including DeepONet \cite{lu2021learning} and Fourier Neural Operator (FNO) \cite{li2020fourier,li2023fourier}---learn mappings between function spaces \cite{kovachki2023neural} and have been applied to accelerate TO \cite{liang2024fourier,yuan2025method}. FNO is attractive because its spectral parameterization enables discretization-invariant learning. Behroozi et al. \cite{behroozi2025sensitivity} introduced Sensitivity-Constrained Neural Operators (SC-NO), demonstrating that co-training neural operators with sensitivity labels ($\partial u / \partial p$ for scalar physical parameters $p$) improves prediction accuracy and robustness for parametric PDEs while keeping the standard FNO architecture unchanged. In this work, we adopt the same sensitivity-constrained training principle for topology optimization and introduce a distinct architectural mechanism---\texttt{forward\_split}---that separates the density channel from context inputs at inference time to obtain deployed sensitivities via automatic differentiation through the predicted objective field. \revgreen{Compared with existing FNO-based TO studies, KATOsuper differs in how the learned operator enters the optimization loop. Ref.~\cite{liang2024fourier} uses a one-shot topology-prediction setting without in-loop sensitivity updates, while Ref.~\cite{yuan2025method} uses a forward field surrogate within a conventional SIMP loop whose sensitivities are still obtained through the classical optimization formulation. In contrast, SC-FNO predicts the objective field and obtains the deployed density sensitivity by automatic differentiation through the same prediction path, so the surrogate supplies both the in-loop objective and a gradient-consistent update direction.}}

\rev{However, existing surrogate approaches possess several limitations. Primary issues include gradient inconsistency---most surrogates predict the objective $J$ and sensitivity $S = \partial J / \partial \rho$ through separate outputs \cite{liang2024fourier}, violating the Taylor expansion $J(\rho + \delta) \approx J(\rho) + S \cdot \delta$ and producing inconsistent gradients that accumulate errors over iterations. Additionally, generalization limitations arise as convolutional surrogates require retraining for new resolutions, and domain shift occurs in TO when generators explore configurations unseen during training. Furthermore, black-box solver incompatibility presents challenges since commercial FEA packages often lack adjoint access, yet gradient-free methods \cite{kus2024gradient,kato2023tackling} sacrifice convergence quality. Finally, deployment barriers remain because many high-performance frameworks still depend on Linux-centric toolchains and MPI workflows.}

\rev{This work addresses these gaps through KATOsuper, an objective-agnostic framework that extends neural-reparameterized TO with surrogate acceleration and zero-shot generalization. KATOsuper integrates a sensitivity-consistent Fourier neural operator (SC-FNO) that replaces inner-loop FEA with learned predictions while ensuring that the deployed sensitivity is obtained through the same objective-prediction path used during inference. The key insight demonstrated in this study is that sensitivity should be derived via automatic differentiation from the objective prediction rather than predicted independently, preserving consistency between the predicted objective field and the gradient used by the optimizer.}

\begin{figure}[pos=h]
  \centering
  \includegraphics[width=1.0\columnwidth]{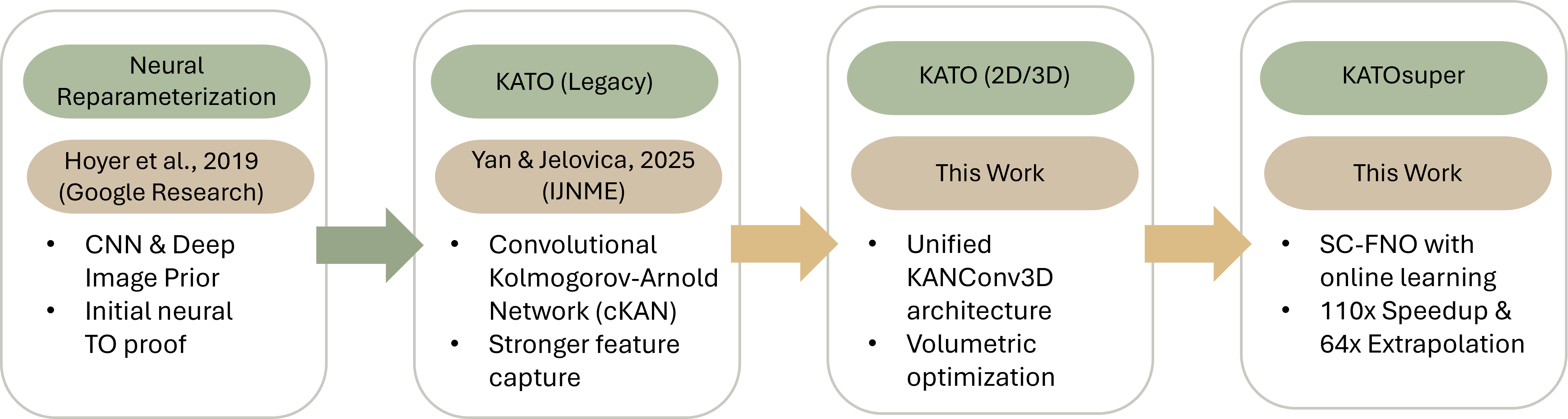}
  \caption{Roadmap of the KATO framework from basic neural reparameterization to the surrogate-accelerated KATOsuper model.}
  \label{fig:roadmap}
\end{figure}

\rev{The KATO framework has evolved from foundational reparameterization to high-performance surrogate-accelerated systems, as illustrated in Fig. \ref{fig:roadmap}. The sequence begins with Neural Reparameterization \cite{hoyer2019neural}, which established the foundation of neural TO using CNNs. This was followed by \revgreen{\textbf{KATO(legacy)}} \cite{yan2025kato}, which enhanced structural integrity through Kolmogorov-Arnold Network (KAN) convolutions to achieve stronger feature capture in stress minimization. In this work, we extend this architecture to \revgreen{\textbf{KATO3D}}, a \revorange{3D neural-reparameterized framework that uses} KANConv3D for volumetric structural optimization. Finally, this progression results in \textbf{KATOsuper}, which integrates a Sensitivity-Consistent Fourier Neural Operator (SC-FNO) that combines the sensitivity-constrained training principle of \cite{behroozi2025sensitivity} with the novel \texttt{forward\_split} inference architecture for surrogate-accelerated optimization, together with optional online learning \cite{hoi2021online}.}

\rev{The input encoding includes Fourier position embedding to enable resolution-invariant learning across 2D and 3D problems. Combined with FNO's spectral architecture and dynamic mode limiting, this enables zero-shot resolution extrapolation beyond the training resolution, including topology-preserving exploration at up to 64 times without retraining. The offline training uses supervised learning with paired objective-sensitivity labels from FEA ground truth, while optional online learning adapts the surrogate using periodic FEA calibration during optimization. The framework supports pure surrogate and hybrid FEA-surrogate modes.}

\rev{This work makes the following contributions: \textbf{KATO3D}, \revorange{a 3D extension of neural-reparameterized topology optimization using a latent-space KANConv3D generator}, tested on compliance and structural stress benchmarks; the \textbf{SC-FNO evaluator} incorporating the \texttt{forward\_split} mechanism---a TO-specific inference-time input decomposition that ensures the deployed sensitivity is obtained as the derivative of the predicted objective field with respect to density. While following the sensitivity-constrained training principle of \cite{behroozi2025sensitivity}, this architectural design is distinct: by separating the density channel (gradient-enabled) from context channels (detached), the same evaluator design can be instantiated for compliance, stress, or other differentiable objectives, providing optimization-ready gradients via automatic differentiation rather than independent regression; the unified \textbf{KATOsuper} framework that delivers \revorange{\textbf{15--110$\times$} deployment-time optimization speedup relative to MATLAB OC/MMA baselines (which are found to be faster than OC/MMA in Python)} while \revgreen{maintaining competitive optimization quality within the baselines considered here}; \textbf{zero-shot resolution extrapolation} beyond the training resolution, where compliance degradation increases progressively with scaling factor although structures remain topologically valid even at 64$\times$; \textbf{online learning} for computationally demanding 3D problems, where periodic FEA calibration improves surrogate fidelity along the encountered trajectory; and a complete implementation with standard dependencies, ensuring accessibility across computing environments.}

Section \ref{method} presents the methodology including SC-FNO architecture, input encoding, and network design. Section \ref{sec:experiments} presents numerical experiments on 2D and 3D benchmark problems, and resolution extrapolation. Section \ref{sec:discussion} discusses implications, limitations and possible future work while Section \ref{sec:conclusion} concludes the paper.

\section{Methodology}\label{method}

\subsection{Problem formulation}
Topology optimization seeks the optimal material distribution within a design domain to minimize an objective function subject to constraints. Following the density-based approach \cite{bendsoe2003topology,sigmund2013topology}, the design domain is discretized into finite elements, each assigned a continuous density variable $\rho_i \in (0,1]$. The optimization problem is:
\begin{equation}\label{eq:general_TO}
\min_{\boldsymbol{\rho}} \; J(\mathbf{u}(\boldsymbol{\rho}), \boldsymbol{\rho}) \quad \text{s.t.} \quad \frac{1}{V_0}\sum_i v_i \rho_i \leq \bar{v}, \quad \mathbf{K}(\boldsymbol{\rho})\mathbf{u} = \mathbf{F}
\end{equation}
where $\boldsymbol{\rho}$ is the vector of element densities, $\mathbf{u}$ is the displacement field, $J$ is the objective function, and $\bar{v}$ is the volume fraction constraint.

The Solid Isotropic Material with Penalization (SIMP) method relates element stiffness to density through a power-law interpolation:
\begin{equation}\label{eq:SIMP}
E_i = E_{\min} + \rho_i^{p_{\text{pen}}} (E_0 - E_{\min})
\end{equation}
where $E_0$ is the base material Young's modulus, $E_{\min} \approx 10^{-9}E_0$ prevents singularity, and $p_{\text{pen}} > 1$ is the penalization power.

For compliance minimization, the objective is:
\begin{equation}\label{eq:compliance}
C(\boldsymbol{\rho}) = \mathbf{F}^T \mathbf{u} = \mathbf{u}^T \mathbf{K}(\boldsymbol{\rho}) \mathbf{u}
\end{equation}

For stress minimization, we adopt the p-norm von Mises stress measure \cite{le2010stress,deng2021efficient}:
\begin{equation}\label{eq:stress_pnorm}
\sigma_{PN} = \left( \frac{1}{N}\sum_{i=1}^{N} \hat{\sigma}_{\text{VM},i}^p \right)^{1/p}, \quad \hat{\sigma}_{\text{VM},i} = \rho_i^q \sigma_{\text{VM},i}
\end{equation}
where $q$ is a relaxation factor and $p$ controls the approximation to maximum stress.

\subsection{KATOsuper architecture overview}
\begin{figure}[pos=h]
  \centering
  \includegraphics[width=0.6\textwidth]{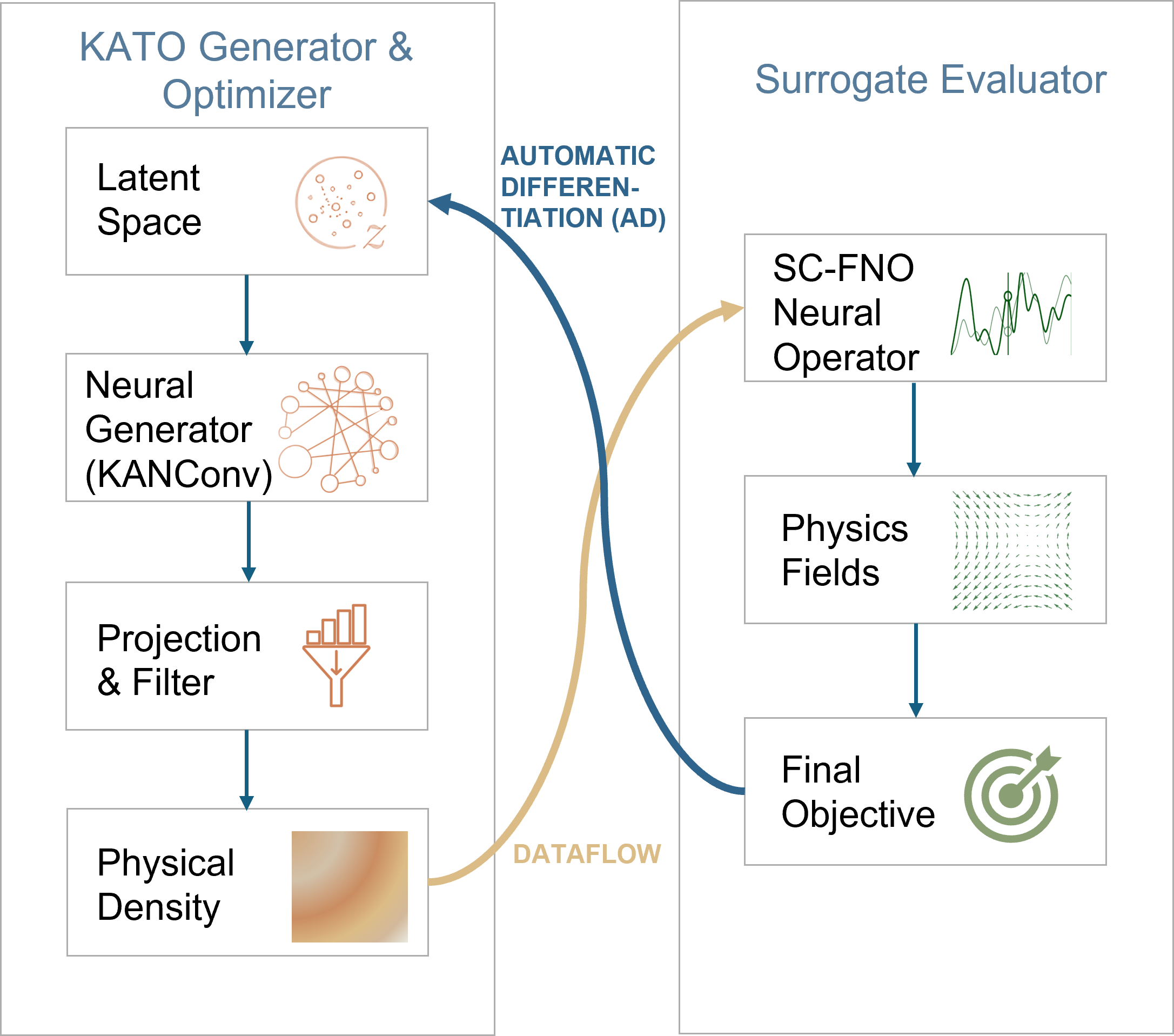}
  \caption{Overview of the KATOsuper Generator-Evaluator framework.}
  \label{fig:architecture}
\end{figure}

KATOsuper decouples design generation from physics evaluation through a Generator-Evaluator framework (Fig.~\ref{fig:architecture}). This modular design enables swapping between FEA and surrogate evaluators without modifying the generator, and forms the foundation for surrogate-accelerated optimization.

The Generator (KATO) is a neural network $G_\theta(\mathbf{z})$ that maps a latent vector $\mathbf{z}$ to a density field $\boldsymbol{\rho}$. Following KATO \cite{yan2025kato}, the generator architecture may use standard convolutional layers or KAN convolutions (KANConv) \cite{bodner2024convolutional,liu2024kan}. A differentiable volume constraint projection ensures the output satisfies the volume fraction constraint $\bar{v}$:
\begin{equation}\label{eq:sigmoid_projection}
\rho_i = \frac{1}{1 + \exp(-(\hat{\rho}_i + \tilde{b}))}
\end{equation}
where $\hat{\rho}_i$ are raw network logits and $\tilde{b}$ is a bias determined via binary search such that the average density $\frac{1}{N}\sum_i \rho_i = \bar{v}$.

\rev{The Evaluator can operate with either the FEA solver or the SC-FNO surrogate. The FEA solver computes the ground-truth objective field and sensitivity $\nabla J / \nabla \boldsymbol{\rho}$ via sparse Cholesky factorization \cite{chen2008algorithm,davis2006direct}, while the SC-FNO surrogate predicts the objective field and derives sensitivity via automatic differentiation. This architecture is \textbf{objective-agnostic} in the sense that the same generator--evaluator formulation extends across differentiable objectives, with the evaluator instantiated for the specific objective of interest. The primary benefit of using a surrogate is computational efficiency: while FEA scales as $O(N^{1.5})$ for sparse direct solvers, SC-FNO inference requires only $O(N \log N)$ operations via FFT.}

The complete optimization loop iterates: generate $\boldsymbol{\rho} = G_\theta(\mathbf{z})$, evaluate $J$ and $\nabla J / \nabla \boldsymbol{\rho}$ via the selected evaluator, and update $\theta$ and $\mathbf{z}$ via gradient descent.

\rev{The choice of optimizer for neural reparameterization has evolved through successive architectural advances. The original neural TO framework \cite{hoyer2019neural} employed standard CNN architectures that exhibited highly non-convex loss landscapes during latent space optimization, necessitating L-BFGS---a quasi-Newton method---to achieve stable convergence. However, the per-iteration cost of L-BFGS scales poorly with parameter count, limiting practical scalability. KATO \cite{yan2025kato} addressed this limitation by replacing fixed-activation CNNs with KAN convolutions featuring learnable B-spline activations. The smoother function approximation provided by B-splines substantially regularizes the loss surface, enabling stable optimization with Adam \cite{kingma2014adam}---a first-order method with orders-of-magnitude lower per-step cost. In KATOsuper, we use AdamW \cite{loshchilov2017decoupled}, whose decoupled weight decay provides more effective regularization of the latent vector $\mathbf{z}$ and generator weights $\theta$ compared to Adam, preventing unbounded latent vector growth and maintaining optimization flexibility. A detailed comparison between optimizers for neural reparameterization was presented in our prior work \cite{yan2025kato}.}

\subsection{Physics embedding: unified input design}
Effective surrogate learning requires encoding not just the density field, but also the complete problem specification including boundary conditions, loads, and material parameters. The SC-FNO surrogate receives a multi-channel input tensor that comprehensively encodes this optimization state and problem geometry. We design a 24-channel encoding for 3D problems as shown in Table~\ref{tab:channels_3d}; for 2D problems, the encoding reduces to 18 channels by removing all $z$-axis related channels (Ch 3, 6, 9, 14, 17, 20).

The encoding primarily utilizes the Fourier position embedding \cite{mildenhall2021nerf,vaswani2017attention}, which maps normalized coordinates to sinusoidal features:
\begin{equation}\label{eq:fourier_pe}
\begin{aligned}
\text{PE}(\mathbf{x}) = [& \sin(\pi x), \sin(\pi y), \sin(\pi z), \\
& \cos(\pi x), \cos(\pi y), \cos(\pi z)]
\end{aligned}
\end{equation}
\rev{Resolution invariance arises from the use of normalized coordinates $\mathbf{x} \in [0,1]^3$: changing from $64^3$ to $128^3$ preserves the same coordinate range, enabling zero-shot resolution transfer. The sinusoidal functions provide spectral richness that helps the FNO distinguish spatial locations within the domain.}

\begin{table}[pos=h]
\centering
\small
\caption{24-channel input encoding for 3D SC-FNO. Channel 0 has gradient tracking. For 2D, remove $z$-axis channels (3, 6, 9, 14, 17, 20) to obtain 18 channels. The encoding is objective-agnostic; $J$ denotes the scalar objective (e.g., compliance or p-norm stress).}
\label{tab:channels_3d}
\begin{tabular}{clp{4cm}}
\hline
Ch. & Content & Purpose \\
\hline
0 & $\rho$ & Density (gradient enabled) \\
1 & $\sin(\pi x)$ & Fourier PE (sine, $x$) \\
2 & $\sin(\pi y)$ & Fourier PE (sine, $y$) \\
3 & $\sin(\pi z)$ & Fourier PE (sine, $z$) \\
4 & $f_x$ & Normalized force ($x$) \\
5 & $f_y$ & Normalized force ($y$) \\
6 & $f_z$ & Normalized force ($z$) \\
7 & BC$_x$ & BC mask ($x$) \\
8 & BC$_y$ & BC mask ($y$) \\
9 & BC$_z$ & BC mask ($z$) \\
10 & $\bar{v}$ & Volume fraction \\
11 & $p_{\text{pen}}$ & SIMP penalization \\
12 & $\cos(\pi x)$ & Fourier PE (cosine, $x$) \\
13 & $\cos(\pi y)$ & Fourier PE (cosine, $y$) \\
14 & $\cos(\pi z)$ & Fourier PE (cosine, $z$) \\
15 & Blur($f_x$) & Blurred force ($x$) \\
16 & Blur($f_y$) & Blurred force ($y$) \\
17 & Blur($f_z$) & Blurred force ($z$) \\
18 & Blur(BC$_x$) & Blurred BC ($x$) \\
19 & Blur(BC$_y$) & Blurred BC ($y$) \\
20 & Blur(BC$_z$) & Blurred BC ($z$) \\
21 & Prior $J$ & Objective prior \\
22 & Prior $\nabla J$ & Sensitivity prior \\
23 & Mask & Design mask \\
\hline
\end{tabular}
\end{table}

Gaussian-blurred fields (channels 15--20) propagate load and boundary condition awareness to distant regions. The blurring is performed via convolution with a Gaussian kernel $G_{\sigma_b}$:
\begin{equation}\label{eq:blur}
\begin{gathered}
F_{\text{blur}} = \mathbf{f} * G_{\sigma_b} \\
G_{\sigma_b}(\mathbf{x}) = \frac{1}{(2\pi\sigma_b^2)^{3/2}} \exp\left(-\frac{\|\mathbf{x}\|^2}{2\sigma_b^2}\right)
\end{gathered}
\end{equation}
with $\sigma_b = 2.0$ voxels, providing smooth spatial gradients indicating proximity to mechanically significant features.

\rev{Behroozi et al. \cite{behroozi2025sensitivity} proposed sensitivity-constrained training for parametric PDEs, where sensitivity labels $\partial u/\partial p$ for scalar parameters $p$ are added as a supervised loss term during training while preserving the standard FNO architecture. Our work follows the same training principle---the offline loss includes a sensitivity supervision term---but introduces a TO-specific architectural modification absent from \cite{behroozi2025sensitivity}. The \texttt{forward\_split} mechanism separates the density channel from context channels at inference, creating a dedicated gradient path from the predicted objective field to density inputs. This ensures that the deployed sensitivity is obtained as $\partial J_{\mathrm{pred}}/\partial \rho$ through automatic differentiation, rather than being predicted by an independent output head. The operator maps density fields to objective fields, requiring the comprehensive physics encoding of Table~\ref{tab:channels_3d}, which is fundamentally different from the scalar-parameter inputs of \cite{behroozi2025sensitivity}.}

\rev{The SC-FNO architecture incorporates the \texttt{forward\_split} mechanism that ensures physically consistent sensitivities for any differentiable objective. During the forward pass, the input is split into the density channel $\boldsymbol{\rho}$ (with gradient tracking) and context channels $\mathbf{c}$ (detached from the computational graph):}
\begin{equation}\label{eq:forward_split}
\boldsymbol{\rho} \leftarrow \mathbf{x}_{\text{input}}[0], \quad \mathbf{c} \leftarrow \text{detach}(\mathbf{x}_{\text{input}}[1:])
\end{equation}
After the forward pass, sensitivity is computed via automatic differentiation:
\begin{equation}\label{eq:sensitivity_autograd}
S_{\text{pred}} = \frac{\partial}{\partial \boldsymbol{\rho}} \left( \sum_{i,j,k} J_{\text{pred}}^{(i,j,k)} \right)
\end{equation}
\rev{This design ensures that the deployed sensitivity is obtained as the derivative of the predicted objective field with respect to density. For any differentiable objective $J$, the first-order approximation $J(\boldsymbol{\rho} + \boldsymbol{\delta}) \approx J(\boldsymbol{\rho}) + S \cdot \boldsymbol{\delta}$ therefore follows the same computational path used during deployment.}

\subsection{KATO3D: 3D neural reparameterized TO}
\begin{figure}[pos=h]
  \centering
  \includegraphics[width=0.6\textwidth]{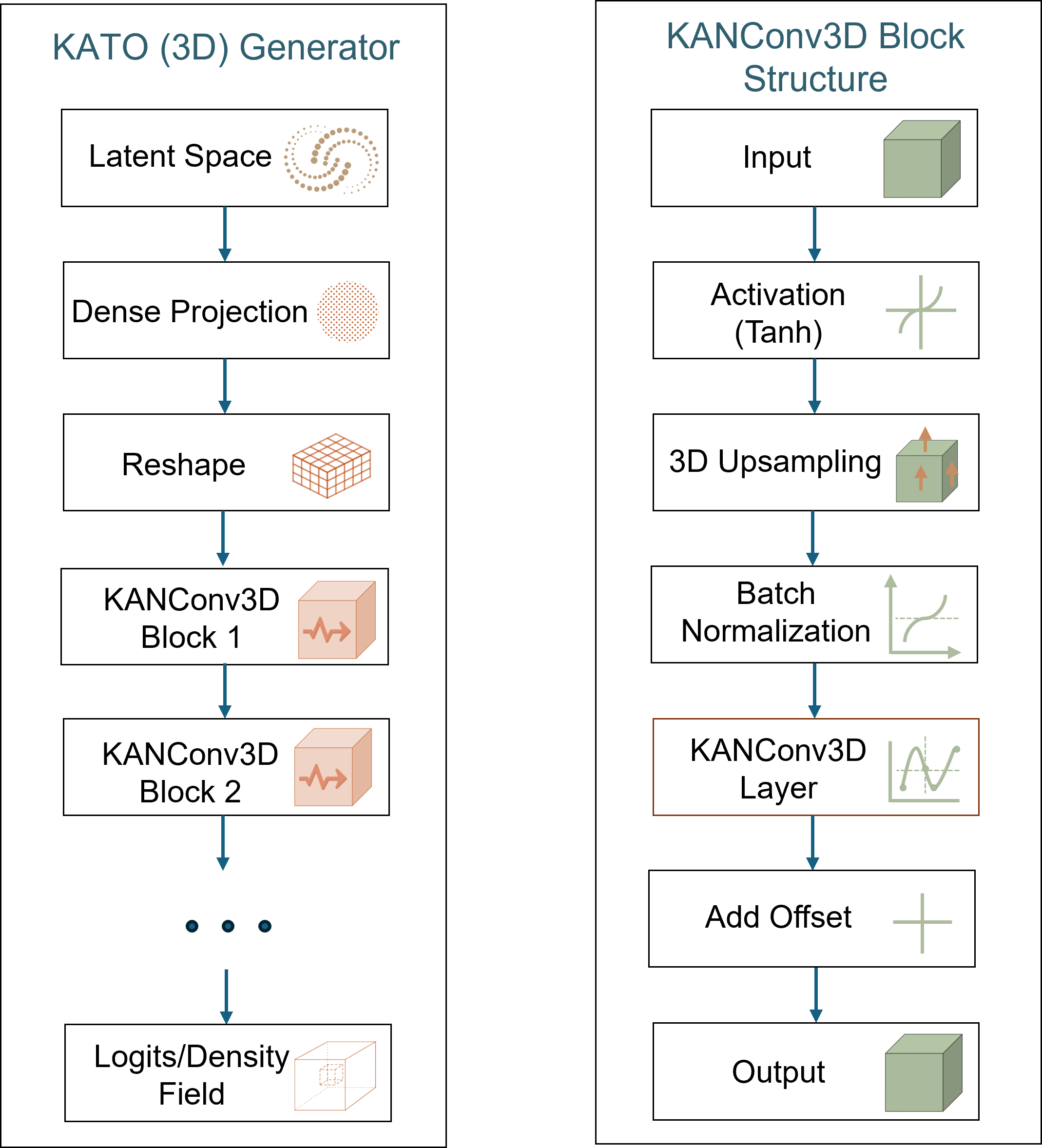}
  \caption{KATO3D generator architecture with stacked KANConv3D blocks.}
  \label{fig:kato3d}
\end{figure}

\rev{We extend the KATO framework to 3D TO through KATO3D, \revorange{a 3D extension of neural-reparameterized topology optimization using a latent-space KANConv3D generator}, as illustrated in Fig. \ref{fig:kato3d}. KATO3D employs a pure KANConv3D generator architecture, leveraging KAN convolutions \cite{bodner2024convolutional,liu2024kan} for enhanced expressiveness. KAN convolutions are employed for their smooth function approximation properties: the learnable B-spline activations act as a continuous nonlinear prior that produces smoother density fields than ReLU-based architectures. A detailed ablation comparing KAN against standard CNN architectures was presented in \cite{yan2025kato}.}

The KANConv3D generator replaces standard convolutional layers with KAN-based convolutions:
\begin{equation}\label{eq:kanconv3d}
\text{KANConv3D}(\mathbf{x}) = \sum_{k=1}^{K} \phi_k(\mathbf{W}_k * \mathbf{x})
\end{equation}
where $\phi_k$ are learnable B-spline activation functions placed on the edges, and $\mathbf{W}_k$ are 3D convolutional kernels. This architecture captures finer topological details through spline-based transformations without increasing network depth.

The generator architecture follows a standardized block design: Tanh activation $\rightarrow$ Trilinear Upsampling $\rightarrow$ LayerNorm $\rightarrow$ KANConv3D + Offset. The network progressively upsamples from a compact latent representation to the full 3D density field.

For 3D FEA, the stiffness matrix $\mathbf{K} \in \mathbb{R}^{3N_{\text{nodes}} \times 3N_{\text{nodes}}}$ is assembled from 8-node hexahedral elements~\cite{zienkiewicz2005finite} using SIMP material interpolation. Element strains are computed from the strain-displacement matrix:
\begin{equation}\label{eq:3d_strain}
\boldsymbol{\varepsilon}_e = \mathbf{B}_e \mathbf{u}_e
\end{equation}
where $\mathbf{B}_e$ is the 6$\times$24 strain-displacement matrix for hexahedral elements. Stresses follow from the 3D constitutive relation $\boldsymbol{\sigma}_e = \mathbf{D}_e \boldsymbol{\varepsilon}_e$, and von Mises stress is aggregated using the p-norm formulation. All operations are vectorized using Einstein summation to avoid explicit element loops.

\subsection{SC-FNO architecture}
\begin{figure}[pos=h]
  \centering
  \includegraphics[width=0.6\textwidth]{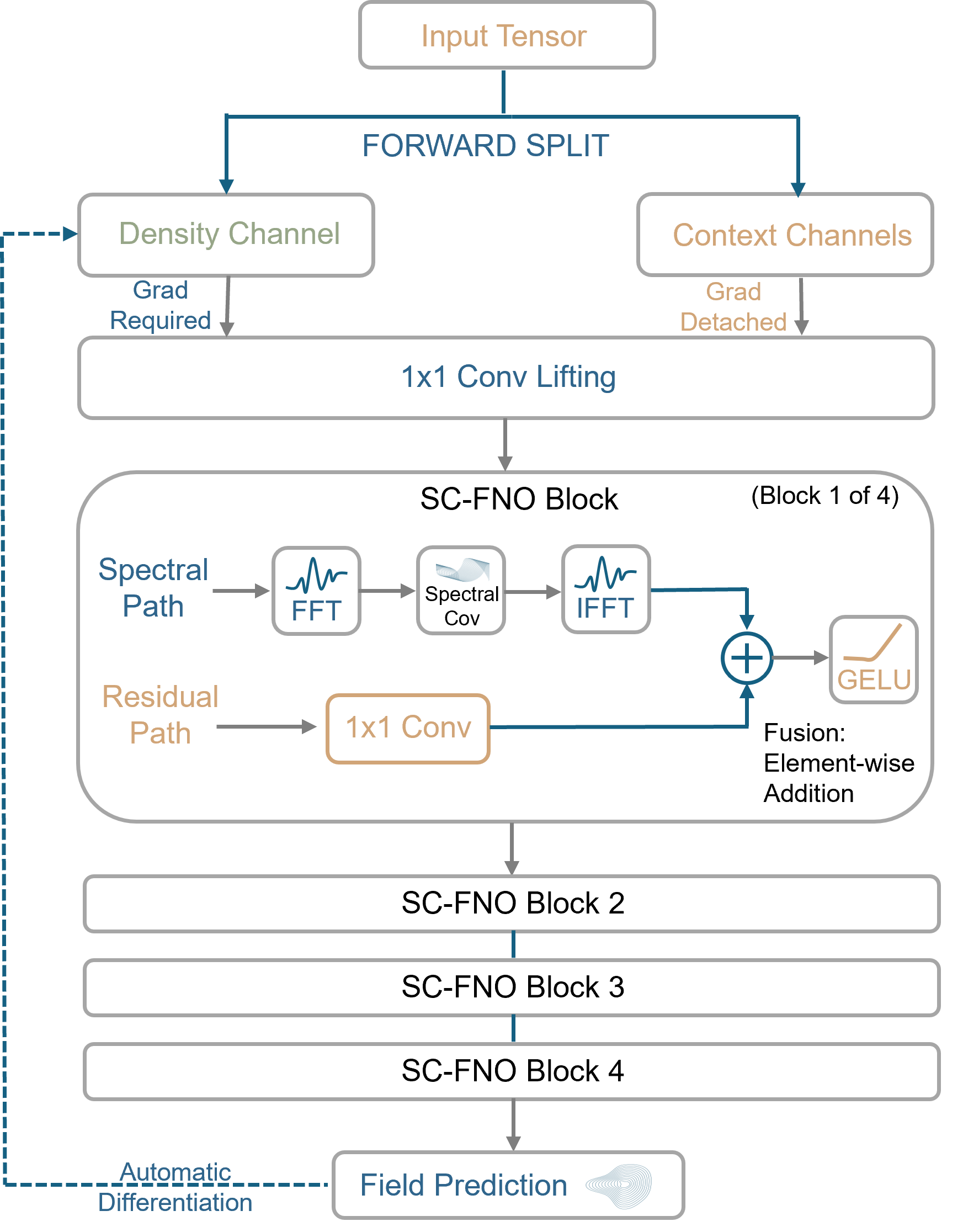}
  \caption{SC-FNO architecture with input projection, spectral convolution layers, and output projection.}
  \label{fig:scfno}
\end{figure}

The Sensitivity-Consistent Fourier Neural Operator (SC-FNO) is designed to predict \textbf{objective fields} \revred{while deriving deployed sensitivities through the same objective-prediction path used at inference}. Fig. \ref{fig:scfno} illustrates the three primary stages of the architecture.

The input projection lifts the multi-channel input to a hidden dimension:
\begin{equation}
\mathbf{h}_0 = \mathbf{W}_{\text{in}} \mathbf{x}_{\text{input}} + \mathbf{b}_{\text{in}}
\end{equation}
where $\mathbf{x}_{\text{input}} \in \mathbb{R}^{C_{in} \times D \times H \times W}$ with $C_{in} = 24$ for 3D and $C_{in} = 18$ for 2D.

Each FNO block applies spectral convolution in the Fourier domain \cite{li2020fourier}:
\begin{equation}\label{eq:fno_layer}
\mathbf{h}_{l+1} = \text{GELU}\left( \mathcal{F}^{-1}\left(\mathbf{R}_l \odot \mathcal{F}(\mathbf{h}_l)\right) + \mathbf{W}_l \mathbf{h}_l \right)
\end{equation}
where $\mathcal{F}$ denotes the FFT, $\mathbf{R}_l \in \mathbb{C}^{d \times d \times m_1 \times m_2 \times m_3}$ are complex-valued spectral weights for the first $m_1 \times m_2 \times m_3$ Fourier modes, $\mathbf{W}_l$ is a 1$\times$1 convolution, and GELU is the Gaussian Error Linear Unit activation \cite{hendrycks2016gaussian}. For 2D, the spectral weights are $\mathbf{R}_l \in \mathbb{C}^{d \times d \times m_1 \times m_2}$.

To enable zero-shot extrapolation to unseen resolutions, the number of retained Fourier modes is dynamically clamped based on the input grid size at inference time:
\begin{equation}\label{eq:mode_limiting}
m'_i = \min(m_i, \lfloor N_i / 2 \rfloor + 1)
\end{equation}
where $N_i$ is the input resolution along dimension $i$. This clamping ensures that at higher resolutions than training, only the modes that fit within the spectral capacity of the trained operator are retained, preventing out-of-bounds spectral access while preserving the learned physics relationships.

The output projection maps back to a single-channel objective field \revred{using the same output-head structure across objective-specific surrogate instances}:
\begin{equation}
J_{\text{pred}} = \mathbf{W}_{\text{out},2} \cdot \text{GELU}(\mathbf{W}_{\text{out},1} \mathbf{h}_L + \mathbf{b}_{\text{out},1}) + \mathbf{b}_{\text{out},2}
\end{equation}
The network outputs the objective field exclusively; sensitivities are derived via automatic differentiation (Eq. \eqref{eq:sensitivity_autograd}), enabling optimization of any differentiable objective \revred{that can be represented through this objective-prediction path} without architecture modification.

\subsection{Data normalization}
Training neural operators on physical quantities requires careful attention to numerical scaling. Objective and sensitivity values in TO span extreme ranges (from $10^{-60}$ in near-void regions to $10^{2}$ in solid regions), which can destabilize gradient-based training. We address this through specialized normalization transforms that are objective-agnostic.

For non-negative objectives (e.g., compliance), we apply the Log1p scaler:
\begin{equation}\label{eq:log1p}
J_{\text{norm}} = \frac{\log(1 + |J|) - \mu}{s}
\end{equation}
where $\mu$ and $s$ are robust statistics (median and interquartile range) from the training distribution.

For \revred{signed or heavy-tailed} objectives (e.g., p-norm stress), we apply the LogModulus scaler:
\begin{equation}\label{eq:logmodulus}
J_{\text{norm}} = \text{sign}(J) \cdot \log(1 + |J|)
\end{equation}
This transform handles both positive and negative values symmetrically.

Log-type transforms provide numerical stability for heavy-tailed distributions \cite{box1964analysis,john1980alternative}. Using median and interquartile range (IQR) as robust location and scale estimators \cite{huber1996robust} reduces sensitivity to outliers common in early optimization stages.

At inference, the chain rule relates normalized to physical quantities:
\begin{equation}\label{eq:jacobian_comp}
\frac{\partial J_{\text{phys}}}{\partial \boldsymbol{\rho}} = s \cdot (J_{\text{phys}} + 1) \cdot \frac{\partial J_{\text{norm}}}{\partial \boldsymbol{\rho}}
\end{equation}
This Jacobian correction ensures physically sound sensitivity during surrogate-accelerated optimization, regardless of the specific objective function.

\subsection{Offline training}
The SC-FNO is pre-trained on trajectory frames extracted from KATO optimization runs. Each frame provides a supervised sample $(\boldsymbol{\rho}, J, \nabla J)$ where the sensitivity is computed via ground-truth FEA. The offline training loss combines three objectives balanced via AWL \cite{kendall2018multi}:
\begin{equation}\label{eq:offline_loss}
\mathcal{L}_{\text{offline}} = \text{AWL}(\mathcal{L}_J, \mathcal{L}_S, \mathcal{L}_{\text{edge}})
\end{equation}
where the individual loss terms are:
\begin{align}
\mathcal{L}_J &= \|J_{\text{pred}} - J_{\text{FEA}}\|_1 \label{eq:loss_comp} \\
\mathcal{L}_S &= \|\nabla J_{\text{pred}} - \nabla J_{\text{FEA}}\|_1 + w_{\text{extreme}} \|\nabla J_{\text{pred}} \cdot M - \nabla J_{\text{FEA}} \cdot M\|_1 \label{eq:loss_sens} \\
\mathcal{L}_{\text{edge}} &= \|\nabla_x \nabla J_{\text{pred}} - \nabla_x \nabla J_{\text{FEA}}\|_1 + \|\nabla_y \nabla J_{\text{pred}} - \nabla_y \nabla J_{\text{FEA}}\|_1 \label{eq:loss_edge}
\end{align}
where $M$ is a mask highlighting extreme sensitivity values (top 5\% most negative), $w_{\text{extreme}} = 5.0$ provides extra weight for critical regions that dominate optimization convergence, and $\nabla_x$, $\nabla_y$ are Sobel gradient operators \cite{sobel1968isotropic} that preserve sharp boundaries in sensitivity predictions.

The AWL mechanism learns optimal weights for each loss term by minimizing $w_i \mathcal{L}_i + \log(\text{var}_i)$, where $w_i = 1/(2\text{var}_i)$ and $\text{var}_i$ are trainable variance parameters. This addresses the heterogeneous scales of objective values (order $10^2$) versus sensitivity magnitudes (order $10^{-60}$ to $10^2$) without manual tuning.
\revgreen{The factor $w_{\mathrm{extreme}}=5$ is applied only to the masked top-5\% sub-term inside $\mathcal{L}_S$, not to the full sensitivity loss; the AWL weights then balance $\mathcal{L}_J$, $\mathcal{L}_S$, and $\mathcal{L}_{\mathrm{edge}}$ at the aggregate loss level.}

\revgreen{For stress objectives, high-magnitude sensitivity regions near loads, supports, and re-entrant corners are handled by three mechanisms. First, \texttt{forward\_split} predicts the smoother objective field and derives sensitivities by automatic differentiation, avoiding direct regression of a singular sensitivity field. Second, the sensitivity loss includes the masked top-5\% extreme-sensitivity term described above, which concentrates training on the regions that dominate stress-driven updates. Third, the Sobel edge loss preserves sharp spatial transitions in the predicted objective field, improving the differentiated sensitivity near geometric discontinuities.}
\revorange{The per-term behavior of the offline loss is evaluated in Sec.~\ref{sec:benchmark_setup}, after the benchmark cases and training configuration are introduced.}

\subsection{Online learning}
\begin{figure}[pos=h]
  \centering
  \includegraphics[width=0.6\textwidth]{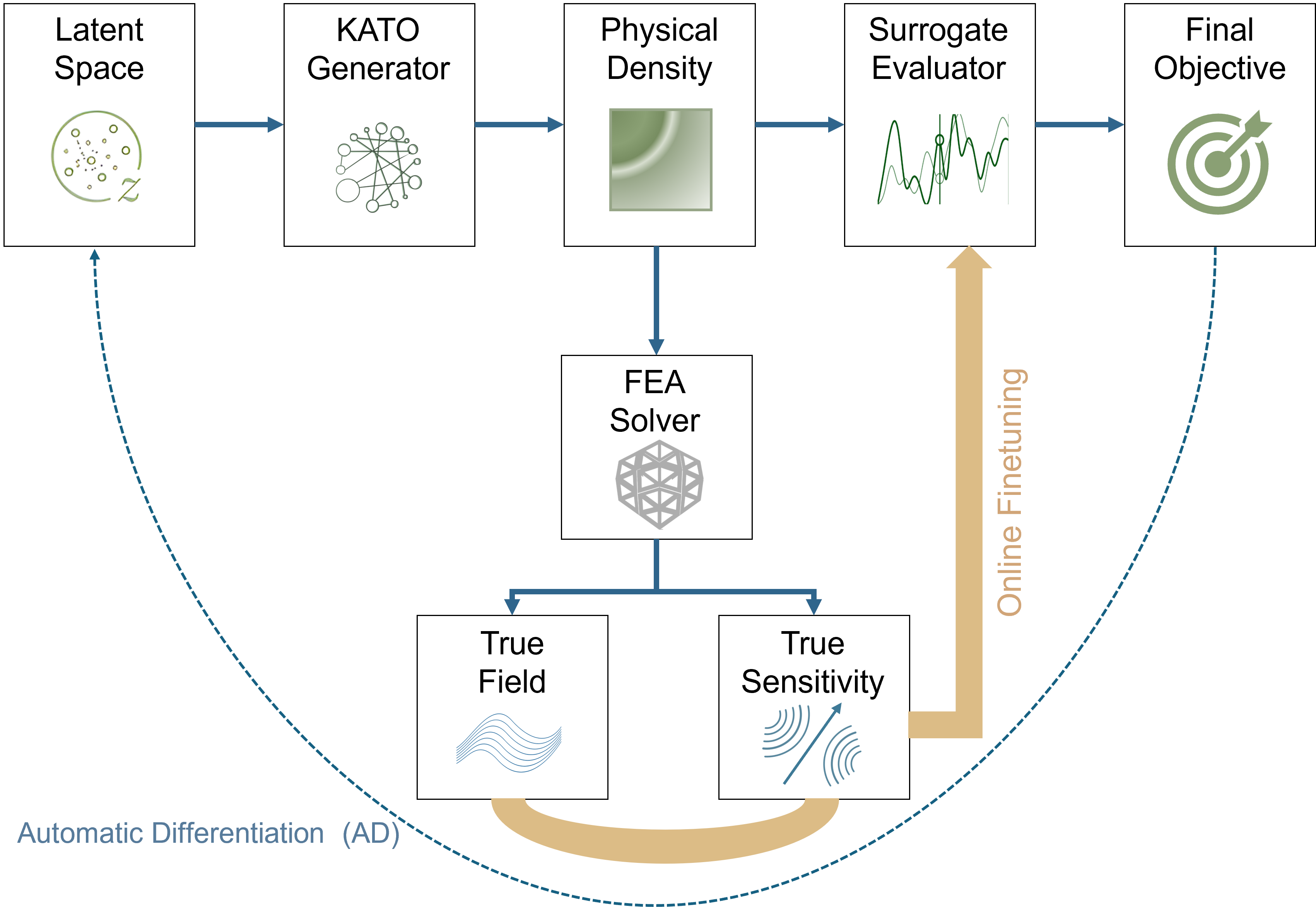}
  \caption{Online learning workflow with periodic FEA calibration.}
  \label{fig:online_learning}
\end{figure}

Surrogates trained offline may fail when the generator explores out-of-distribution (OOD) structural configurations. The online learning strategy (Fig. \ref{fig:online_learning}) adapts the surrogate during optimization using periodic FEA calibration \cite{hoi2021online}.

\rev{The online learning loop proceeds as follows: FEA calibration is performed for the first 5 optimization steps, followed by periodic fine-tuning every $N_{\text{calib}} = 4$ iterations thereafter. This interval was chosen empirically as a practical compromise between calibration frequency and computational overhead; a systematic ablation of this hyperparameter is left for future work. During these steps, the true FEA solver computes ground-truth objective $J_{\text{FEA}}$ and sensitivity $\nabla J_{\text{FEA}}$. The SC-FNO weights $\psi$ are then fine-tuned for $M = 3$ gradient steps before resuming surrogate-accelerated optimization. The online learning loss combines MSE and cosine similarity:}
\begin{equation}\label{eq:online_loss}
\begin{aligned}
\mathcal{L}_{\text{online}} &= \text{MSE}(J_{\text{pred}}, J_{\text{FEA}}) \\
&+ \alpha \text{MSE}(\nabla J_{\text{pred}}, \nabla J_{\text{FEA}}) + \beta \mathcal{L}_{\cos}
\end{aligned}
\end{equation}
\rev{where $\mathcal{L}_{\cos} = 1 - \cos(\nabla J_{\text{pred}}, \nabla J_{\text{FEA}})$ enforces gradient direction alignment. The cosine term is critical because topology optimization is driven by gradient direction, not magnitude. In this setting, every periodic FEA call both validates the current surrogate prediction and provides data for fine-tuning during optimization \cite{hoi2021online}.}
\revgreen{The online loss is intentionally simpler than the offline AWL loss. Offline training uses a large static dataset and can afford objective, sensitivity, and edge reconstruction terms. Online calibration uses a small, shifting batch along the current optimization trajectory, so it prioritizes stable local adaptation through objective MSE, sensitivity MSE, and cosine alignment of the sensitivity direction.}

\subsection{High-performance implementation}
We employ a multi-layered optimization strategy to maximize computational throughput.

\textit{PARDISO Solver:} FEA is accelerated via Intel MKL PARDISO \cite{schenk2004solving} through the \texttt{pypardiso} package, providing 14.6$\times$ speedup over SciPy's \texttt{splu} for sparse Cholesky factorization. A single factorization is computed during the forward pass and reused for adjoint sensitivity computation.

\textit{Multi-Level Caching:} Element-to-global DOF mapping indices (\texttt{edof}) are pre-computed once during mesh initialization and cached for reuse across all iterations, avoiding repeated index assembly. Similarly, matrix sub-sampling indices that identify free (unconstrained) DOFs are cached, eliminating redundant boundary condition filtering during stiffness matrix construction. This achieves 4--8$\times$ speedup for repeated FEA solves within an optimization trajectory.

\textit{Torch Compilation:} PyTorch 2.0's Dynamo compiler \cite{ansel2024pytorch} with \texttt{max-autotune} mode fuses element-wise operations (GELU, LayerNorm, residual addition) into single CUDA kernels, significantly reducing kernel launch overhead for the deep spectral network.

\textit{TF32 Precision:} TensorFloat-32 precision on Ampere+ GPUs provides 3$\times$ throughput gain over FP32 for matrix multiplications in the lifting and projection layers, with sufficient numerical precision for physics regression tasks.

The combined optimizations yield an average speedup of $4$--$6\times$ for FEA solves compared to the default SciPy sparse solver implementation, approaching the performance of optimized MATLAB solvers.

\vspace{1\baselineskip}

\textit{Code availability:}
The complete KATOsuper implementation, including the generator and the SC-FNO surrogate, is publicly
available at \url{https://github.com/ysyysy115/KATOsuper}.

\section{Experiments}\label{sec:experiments}

\subsection{\revorange{Benchmark cases and experimental setup}}\label{sec:benchmark_setup}
\begin{figure}[pos=h]
  \centering
  \includegraphics[width=0.8\textwidth]{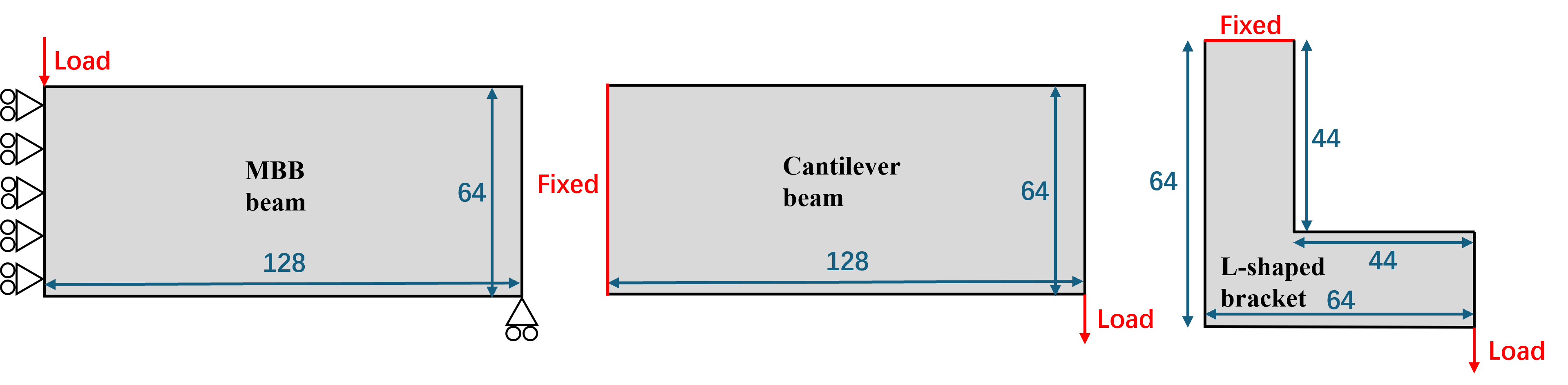}
  \caption{Benchmark design domains: cantilever beam, MBB beam, and L-shaped bracket with their boundary conditions and load configurations.}
  \label{fig:design_domains}
\end{figure}

Table~\ref{tab:method_mapping} summarizes the numerical experiments used to validate the KATOsuper framework across objectives, geometries, and resolutions. \revorange{The 2D cases use the three benchmark geometries shown in Fig.~\ref{fig:design_domains}: MBB beam, cantilever beam, and L-shaped bracket. The cantilever and MBB cases use $128\times64$ elements, while the L-shaped bracket uses $128\times128$ elements. The same geometry family is used for compliance minimization and p-norm stress minimization; the latter uses the physical loading and material parameters stated below.}
\revorange{The 3D experiments use three cases. The standard 3D cantilever ($64\times32\times4$) tests the volumetric extension on a compact cantilever geometry. The 3D split-load cantilever ($32\times32\times16$) applies two symmetric $5\times5$ downward load patches at the front and back of the free end, requiring load transfer through the domain depth. The ship seat-base ($64\times32\times64$) is included as a supplementary larger-geometry check. For the 3D cases, the boundary and load indicators are shown in the leftmost column or panel of the corresponding result figures for brevity, rather than in a separate setup figure.}

For stress optimization experiments, we use realistic physical parameters: applied force $F = 50$ kN, out-of-plane thickness $t = 10$ mm, domain dimensions $1280 \times 640$ mm (corresponding to $128 \times 64$ elements), and steel material properties ($E = 210$ GPa, $\nu = 0.3$) under linear elastic assumptions. The resulting von Mises stresses are reported in MPa. For compliance experiments, we use nominal (unit) material properties to maintain tractable compliance values.

The SC-FNO surrogate is trained on optimization trajectory \textit{frames}---snapshots of the density field and corresponding physics responses extracted from complete KATO optimization runs. Each frame provides a training sample $({\boldsymbol{\rho}, J, \nabla J})$. We generate diverse trajectories by varying:
\begin{itemize}
    \item \textbf{Scenarios}: Multiple boundary condition and load configurations based on Hoyer et al.'s research~\cite{hoyer2019neural}, including cantilever, MBB, and L-shape benchmark problems.
    \item \textbf{\revorange{Initializations}}: 5-10 random initializations per scenario to capture frame diversity.
    \item \textbf{Volume fraction}: Randomly sampled $\bar{v} \in [0.25, 0.40]$ per run.
\end{itemize}

From each run (50 steps for compliance, 100 steps for stress), we extract 40 frames using a hybrid sampling policy: 20 frames from early iterations (capturing rapid topology emergence), 5 frames from final iterations (capturing converged boundaries), and 15 frames via greedy sampling (maximizing density matrix diversity). \revgreen{For the greedy subset, each added middle frame is chosen as the candidate whose maximum cosine similarity to the already selected density fields is smallest, i.e., a farthest-point rule in density space.} \rev{This yields approximately 6,200 frames for 2D compliance training and 3,200 frames for 3D cases (Table~\ref{tab:config}).} \revorange{The 2D surrogates are trained on a broader set of benchmark trajectories covering the same geometry families used in the 2D compliance and stress tests; therefore, the main 2D benchmark evaluations are in-distribution with respect to geometry and boundary-condition type, while the higher-resolution cantilever cases test resolution extrapolation. In contrast, the 3D SC-FNO is trained only on the standard $64\times32\times4$ cantilever trajectories. The 3D split-load cantilever and ship seat-base therefore test out-of-distribution boundary-condition/geometry transfer, and are evaluated with online FEA calibration rather than as purely offline surrogate predictions.}

The complete experimental setup involves three components: the experimental design matrix (Table~\ref{tab:method_mapping}), surrogate training configurations (Table~\ref{tab:config}), and TO parameters (Table~\ref{tab:to_params}).

\begin{table}[h]
\centering
\small
\caption{Overview of numerical experiments across benchmarks.}
\label{tab:method_mapping}
\begin{tabular}{lllll}
\hline
Experiment & Geometry & Objective & Resolution & Methods \\
\hline
\multicolumn{5}{l}{\textit{2D Compliance Optimization}} \\
& Cantilever & Compliance & $128 \times 64$ & OC, KATO, KATOsuper \\
& MBB & Compliance & $128 \times 64$ & OC, KATO, KATOsuper \\
& L-bracket & Compliance & $128 \times 128$ & OC, KATO, KATOsuper \\
\hline
\multicolumn{5}{l}{\textit{2D Stress Optimization}} \\
& Cantilever & P-norm Stress & $128 \times 64$ & MMA, KATO, KATOsuper \\
& MBB & P-norm Stress & $128 \times 64$ & MMA, KATO, KATOsuper \\
& L-bracket & P-norm Stress & $128 \times 128$ & MMA, KATO, KATOsuper \\
\hline
\multicolumn{5}{l}{\textit{Resolution Extrapolation}} \\
& Cantilever & Compliance & $256 \times 128$ (4$\times$) & OC, KATO, KATOsuper \\
& Cantilever & Compliance & $512 \times 256$ (16$\times$) & KATO, KATOsuper \\
& Cantilever & Compliance & $1024 \times 512$ (64$\times$) & OC, KATO, KATOsuper \\
\hline
\multicolumn{5}{l}{\textit{3D Extension}} \\
& \revorange{Standard 3D cantilever} & Compliance & $64 \times 32 \times 4$ & OC, KATO, KATOsuper (Online) \\
& \revorange{3D split-load cantilever} & Compliance & $32 \times 32 \times 16$ & OC, KATO, KATOsuper (Online) \\
& \revorange{Ship seat-base} & Compliance & $64 \times 32 \times 64$ & \revorange{KATO, KATOsuper (Online)} \\
\hline
\end{tabular}
\end{table}

\begin{table}[h]
\centering
\small
\caption{\rev{\revred{SC-FNO training configurations for 2D and 3D compliance surrogates.}}}
\label{tab:config}
\begin{tabular}{lll}
\hline
Parameter & 2D & 3D \\
\hline
Input channels & 18 & 24 \\
Spectral layers & 5 & 4 \\
Hidden width & 96 & 48 \\
Fourier modes & \rev{28} & $(2, 16, 32)$ for $(z, y, x)$ \\
Optimizer & AdamW & AdamW \\
Learning rate & $1 \times 10^{-3}$ (cosine decay) & $1 \times 10^{-3}$ (cosine decay) \\
Batch size & 12 & 8 \\
Epochs & 200 & 100 \\
\rev{Training data runs} & \rev{$\sim$154} & \rev{$\sim$80} \\
Training samples & \rev{$\sim$6,200 frames} & \rev{$\sim$3,200 frames} \\
\rev{Data generation time} & \rev{$\sim$0.5\,h (CPU)} & \rev{$\sim$1.1\,h (CPU)} \\
\rev{SC-FNO training time} & \rev{$\sim$4.8\,h (GPU)} & \rev{$\sim$1.9\,h (GPU)} \\
\hline
\end{tabular}
\end{table}

\begin{figure}[pos=h]
  \centering
  \includegraphics[width=0.72\textwidth]{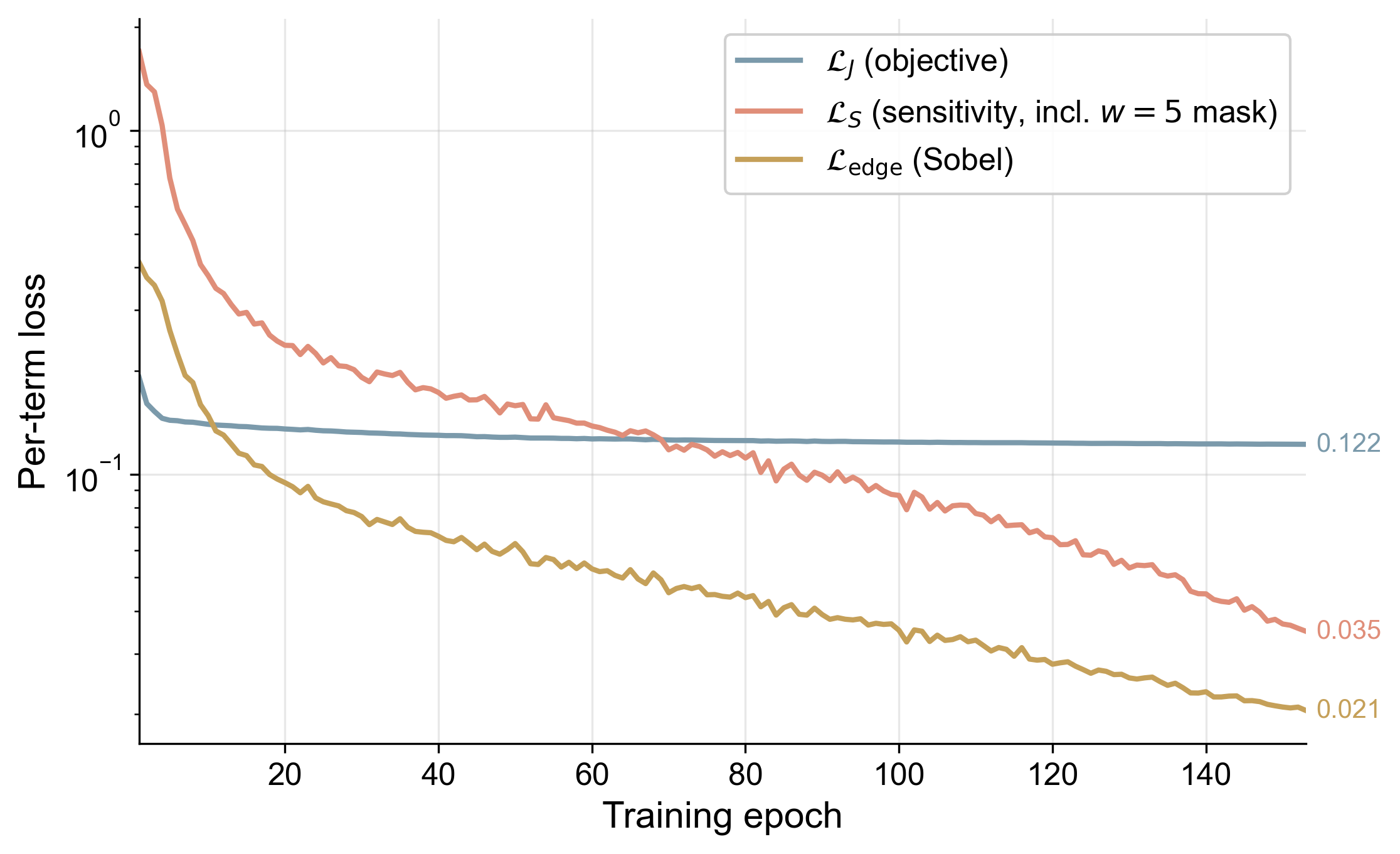}
  \caption{\revorange{Offline training loss components for the 2D compliance SC-FNO.}}
  \label{fig:awl_loss_curves}
\end{figure}

\revorange{As a training diagnostic, Fig.~\ref{fig:awl_loss_curves} reports the per-term loss evolution for the 2D compliance SC-FNO before AWL weighting. The sensitivity loss is initially the largest term because sensitivity fields contain sharper local extrema than objective fields, and the top-5\% mask deliberately emphasizes the most critical high-gradient regions. Its decrease from about 1.7 to 0.035 shows that the masked sensitivity term remains trainable rather than destabilizing the optimization. The Sobel edge loss also decreases steadily to 0.021, indicating stable edge-preservation learning. In contrast, the objective-field loss starts from a lower value and decreases more mildly to 0.122, as expected for the smoother compliance field. These trends show that the masked sensitivity weighting improves attention to critical regions without dominating the AWL-balanced offline objective.}

\begin{table}[h]
\centering
\small
\caption{Topology optimization parameters for compliance and stress benchmarks.}
\label{tab:to_params}
\begin{tabular*}{\columnwidth}{@{\extracolsep{\fill}}lll@{}}
\hline
Parameter & Compliance TO & Stress TO \\
\hline
SIMP penalization ($p_{\text{pen}}$) & 3.0 & 3.0 \\
Minimum density ($\rho_{\min}$) & $10^{-3}$ & $10^{-3}$ \\
Filter radius ($R$) & \multicolumn{2}{l}{$\max(1.5, 0.05 \cdot \max(N_x, N_y, N_z))$} \\
Optimizer (\revorange{Baseline}) & OC & MMA \\
Optimizer (KATO) & AdamW & AdamW \\
Max. iterations & 50 & 100 \\
Volume fraction ($\bar{v}$) & 0.30 & 0.40 \\
Objective function & Total compliance & $p$-norm von Mises stress \\
Stress relaxation factor ($q$) & N/A & 0.5 \\
$p$-norm power ($p$) & N/A & 6.0 \\
\hline
\end{tabular*}
\end{table}

Based on the experimental design in Table~\ref{tab:method_mapping}, we compare three methods: (1) \textbf{OC/MMA (MATLAB)}: conventional optimality criteria or MMA algorithms; (2) \textbf{KATO}: neural-reparameterized optimization with \revorange{direct FEA evaluation}; and (3) \textbf{KATOsuper}: surrogate-accelerated optimization using SC-FNO. Compliance optimization is run for 50 iterations with volume fraction $\bar{v} = 0.30$ following classical TO benchmarks \cite{sigmund200199,liu2014efficient}, while stress optimization is run for 100 iterations with $\bar{v} = 0.40$ following KATO \cite{yan2025kato} to provide sufficient material for stress objective satisfaction. The different configurations also allow evaluation of KATOsuper's generalization across varying iteration counts and material budgets. \revgreen{The experiments were executed on a workstation equipped with an Intel Core i7-14700F CPU (20 cores, up to 5.40~GHz) with dual-channel 32~GB DDR5-5600 RAM for FEA, and an NVIDIA GeForce RTX 4070 GPU (Ada Lovelace architecture, 5,888 CUDA cores, with a base/boost clock of 1,920/2,475~MHz) featuring 12~GB of GDDR6X memory running at 21~Gbps for surrogate training and inference.}

\revred{The MATLAB OC/MMA are used as conventional academic baselines using efficient implementations~\cite{liu2014efficient,deng2021efficient}: they follow standard TO benchmarks and benefit from MATLAB's mature sparse linear-algebra routines. To separate the effect of surrogate acceleration from language and hardware differences, we also report KATO with direct Python-based FEA on the same CPU. The resulting comparisons should therefore be interpreted in two ways: OC/MMA provides the traditional benchmark reference, while KATO provides the same-hardware FEA baseline for isolating the gain from replacing repeated FEA calls with SC-FNO inference.}
\revgreen{For the efficiency comparison, we additionally include a Python implementation of the same OC/MMA baselines using open-source SciPy sparse linear algebra. This same-language benchmark separates the acceleration due to SC-FNO from differences between MATLAB and Python implementations.}

\subsection{2D compliance optimization}
\begin{figure}[pos=h]
  \centering
  \includegraphics[width=0.6\textwidth]{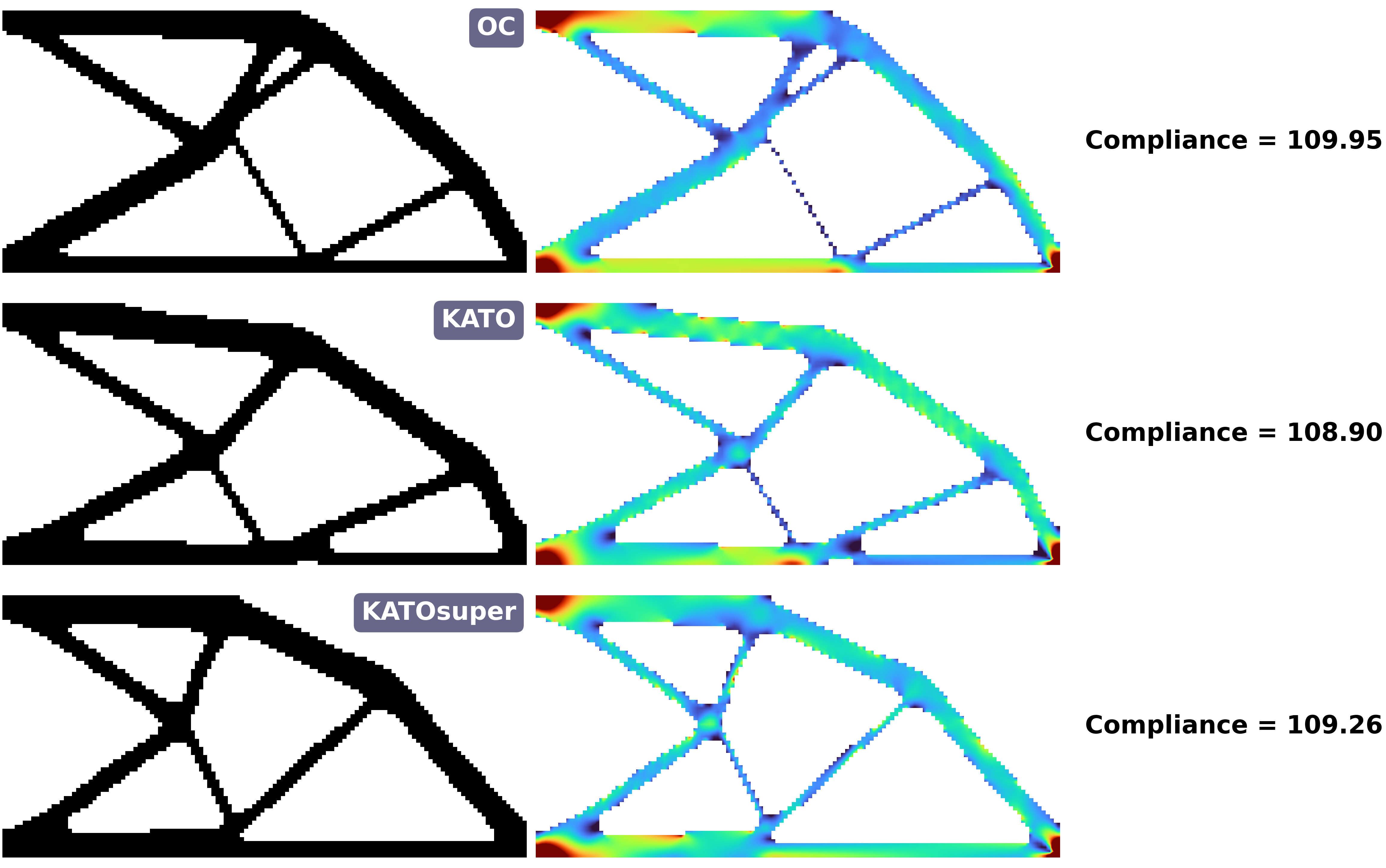}
  \caption{Cantilever beam compliance optimization results at $128 \times 64$ resolution: OC (top), KATO (middle), and KATOsuper (bottom).}
  \label{fig:compliance_2d}
\end{figure}

\begin{figure}[pos=h]
  \centering
  \includegraphics[width=0.6\textwidth]{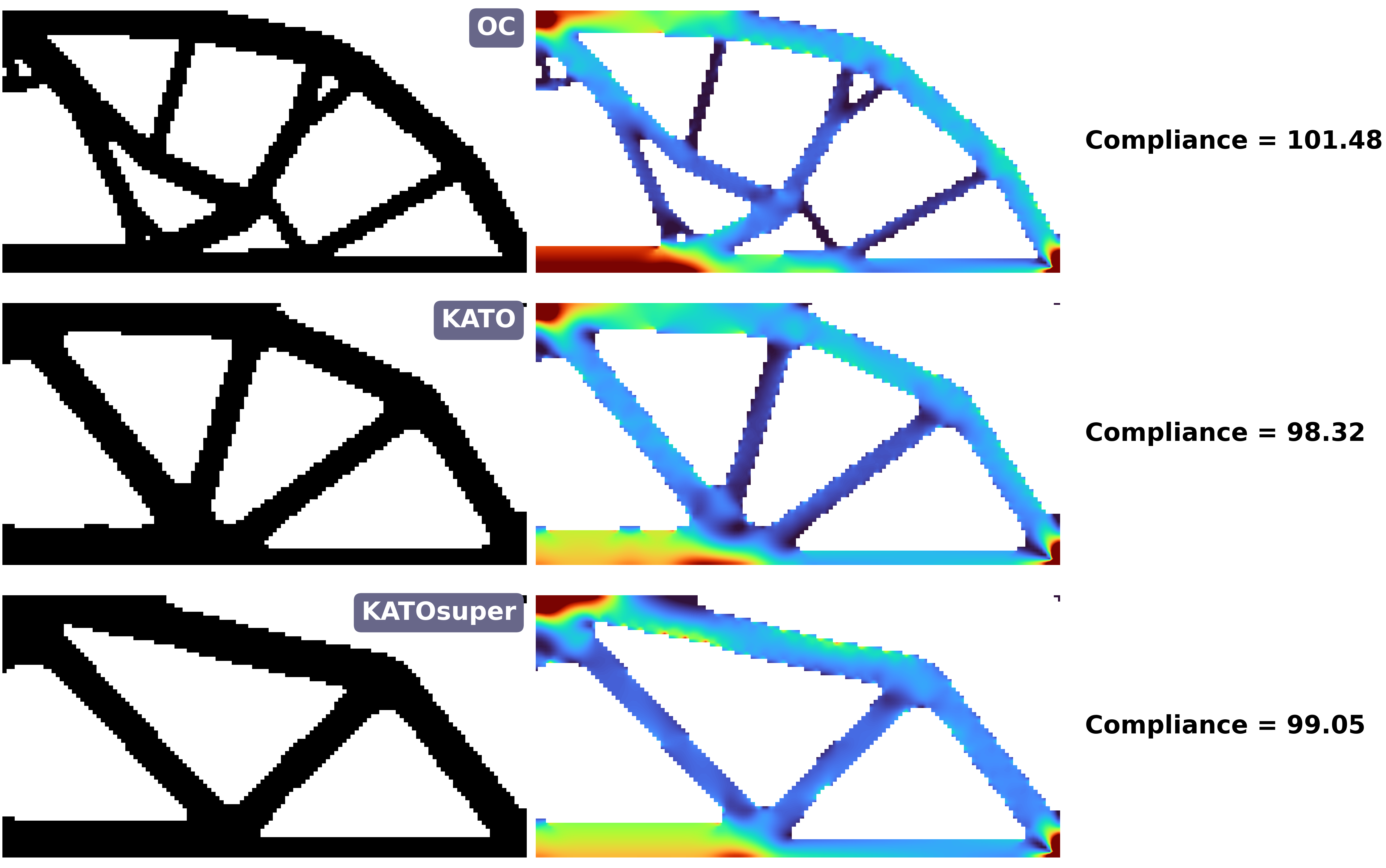}
  \caption{MBB beam compliance optimization results at $128 \times 64$ resolution.}
  \label{fig:compliance_mbb}
\end{figure}

\begin{figure}[pos=h]
  \centering
  \includegraphics[width=0.6\textwidth]{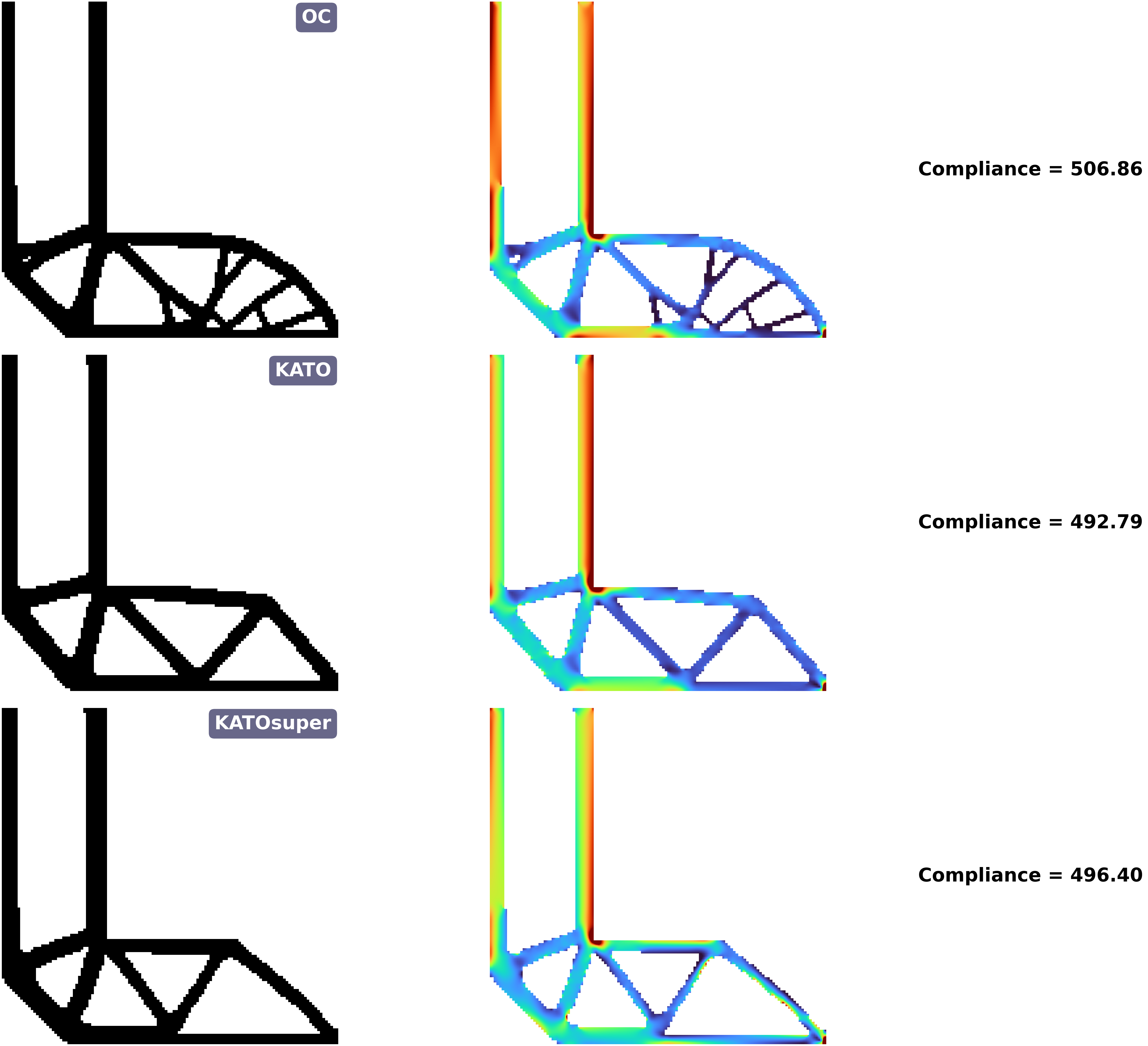}
  \caption{L-shaped bracket compliance optimization results at $128 \times 128$ resolution.}
  \label{fig:compliance_lshape}
\end{figure}

We evaluate compliance minimization on three canonical benchmarks. Figs. \ref{fig:compliance_2d}--\ref{fig:compliance_lshape} present the optimized structures with annotated compliance values.

For the cantilever beam (Fig.~\ref{fig:compliance_2d}), OC achieves compliance of 109.95, while KATO obtains 108.90 and KATOsuper reaches 109.26. The neural methods demonstrate competitive performance, with both KATO and KATOsuper achieving a slight improvement over OC. The MBB beam (Fig.~\ref{fig:compliance_mbb}) also shows KATO's advantage: compliance of 98.32 outperforms both OC (101.48) and KATOsuper (99.05), demonstrating the neural generator's ability to discover superior load paths on symmetric problems. For the geometrically complex L-shaped bracket (Fig.~\ref{fig:compliance_lshape}), KATO again achieves 492.79 \revgreen{and KATOsuper reaches 496.40}, versus OC's 506.86---a 2.8\% \revgreen{KATO} improvement \revgreen{over OC} that highlights the advantage of global optimization on challenging geometries.

\revgreen{KATO gives the lowest compliance on the MBB and L-shape benchmarks in this set}, indicating that neural reparameterization effectively captures the low-dimensional manifold of optimal topologies. Unlike conventional element-by-element optimization, the neural generator learns global relationships between structural members, enabling discovery of well-connected load paths with fewer local minima. The smooth mapping from latent space to density field provides implicit regularization that promotes binary structures without requiring explicit projection filters. \revgreen{We do not interpret these finite nonconvex benchmarks as a guarantee of uniform dominance over OC; rather, they indicate that the neural generator can be advantageous on geometries where globally connected load paths are important.}

Examining the structural differences more closely, KATO produces topologies characterized by smoother boundaries and enhanced structural connectivity compared to traditional OC. This result stems from the generator architecture: the neural network mapping constrains the design to a space of continuous, well-defined geometries, inherently suppressing numerical artifacts such as checkerboard patterns and isolated material islands. From a manufacturing perspective, this implicit regularization ensures high geometric integrity in the final binarized designs, significantly reducing the need for post-processing or manual refinement prior to fabrication.

KATOsuper closely tracks KATO's performance, with differences in compliance below 1\% on all the \revgreen{2D compliance} benchmarks. These results suggest that SC-FNO has learned an accurate mapping from density to compliance. More importantly, it demonstrates that the surrogate provides \textit{directionally correct} sensitivity estimates: even if the predicted compliance magnitude contains small errors, the gradient direction guides the optimizer toward topology improvements. This is a fundamental advantage of the KATOsuper architecture---the neural generator can compensate for approximate sensitivity signals through iterative gradient descent, progressively refining the topology as long as the sensitivity direction remains correct. This contrasts with approaches that directly predict final topologies or attempt to regress sensitivity fields without gradient consistency guarantees, which lack such error-correction mechanisms.

\begin{figure}[pos=h]
  \centering
  \includegraphics[width=0.6\textwidth]{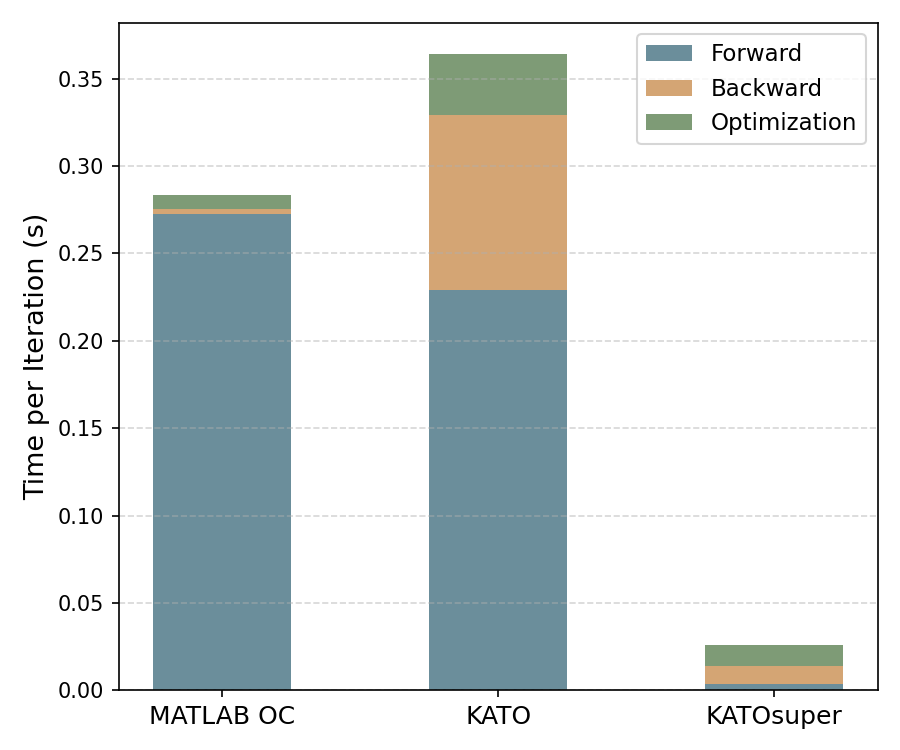}
  \caption{Computational time breakdown for 2D compliance optimization (Cantilever $128 \times 64$).}
  \label{fig:time_compliance}
\end{figure}

The small compliance gaps in KATOsuper results stem primarily from sensitivity prediction errors during early optimization stages, when the density field is still far from the training distribution (near-uniform density versus converged topology). As optimization progresses and the structure approaches the training manifold, surrogate accuracy improves and the final topologies become nearly indistinguishable from KATO results. This observation suggests that adaptive refinement strategies---using surrogate in early stages and switching to FEA for final iterations---could combine the speed of KATOsuper with the precision of KATO. Fig.~\ref{fig:time_compliance} decomposes computation time by operation for the cantilever benchmark. For MATLAB OC, each iteration requires approximately 0.28s, dominated by forward FEA solving (0.26s). KATO achieves similar per-iteration time (0.37s) because it still relies on CPU-based FEA, with additional overhead from generator inference. KATOsuper, by contrast, completes each iteration in just 0.025s by replacing the FEA solver with GPU-accelerated SC-FNO inference. The forward surrogate evaluation accounts for the majority of this time, while backward differentiation and optimization steps remain negligible. This results in a \textbf{19$\times$ speedup} over MATLAB OC at $128 \times 64$ resolution for the whole optimization run, with the acceleration factor increasing at higher resolutions due to the favorable $O(N \log N)$ scaling of spectral operators.

\subsection{2D stress optimization}
\begin{figure}[pos=h]
  \centering
  \includegraphics[width=0.6\textwidth]{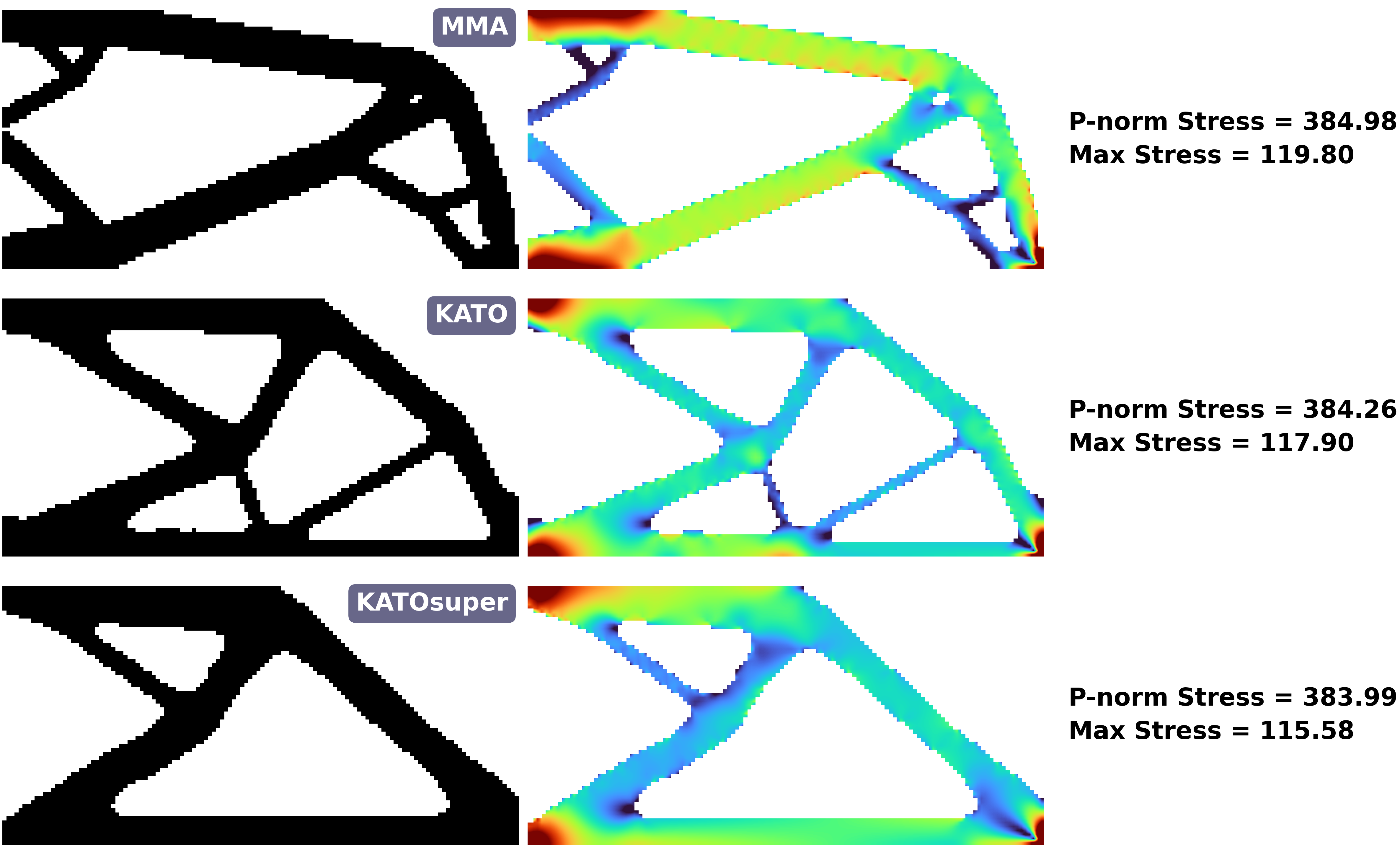}
  \caption{Cantilever beam stress optimization results showing p-norm and maximum von Mises stress.}
  \label{fig:stress_2d}
\end{figure}

\begin{figure}[pos=h]
  \centering
  \includegraphics[width=0.6\textwidth]{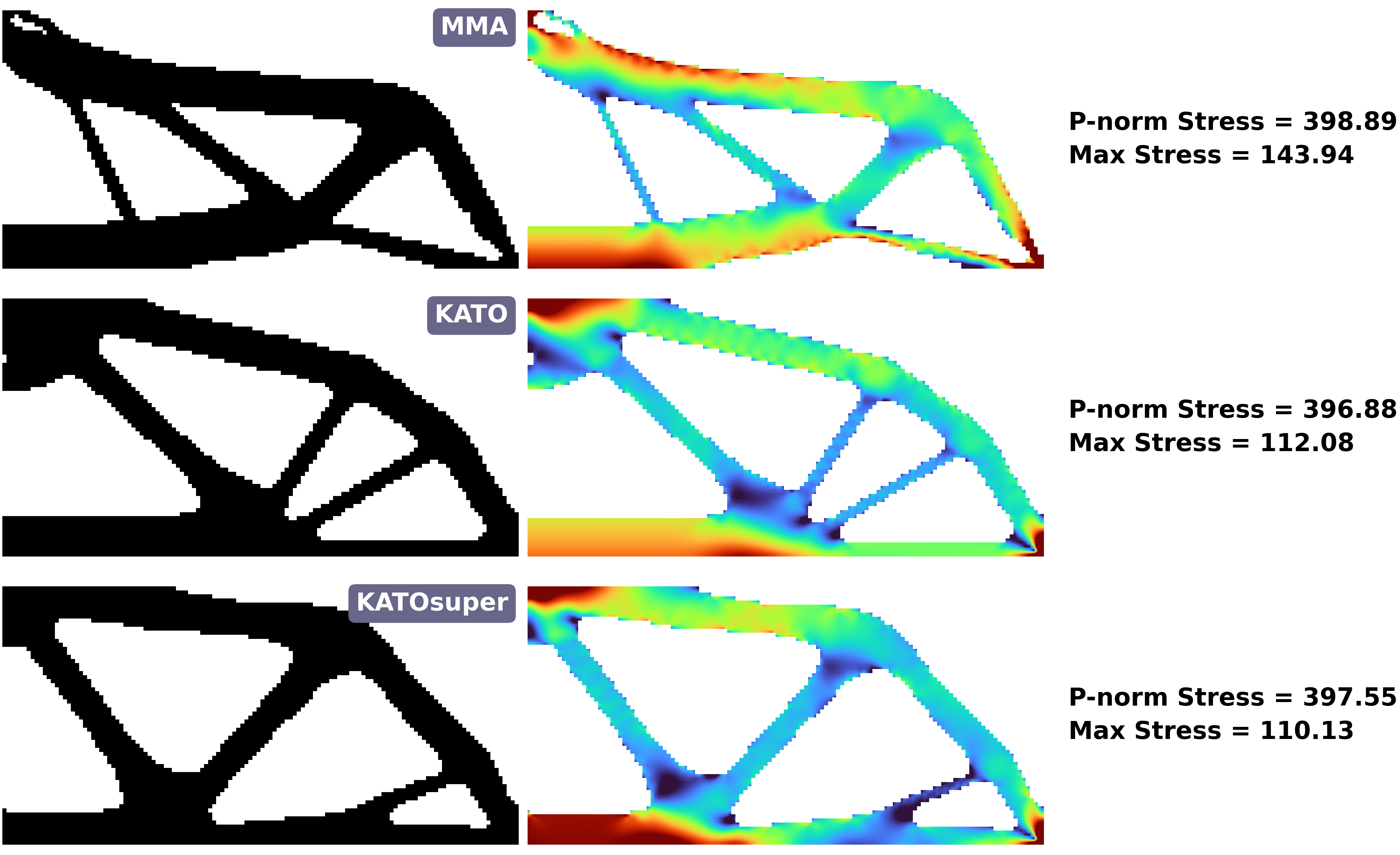}
  \caption{MBB beam stress optimization results.}
  \label{fig:stress_mbb}
\end{figure}

\begin{figure}[pos=h]
  \centering
  \includegraphics[width=0.6\textwidth]{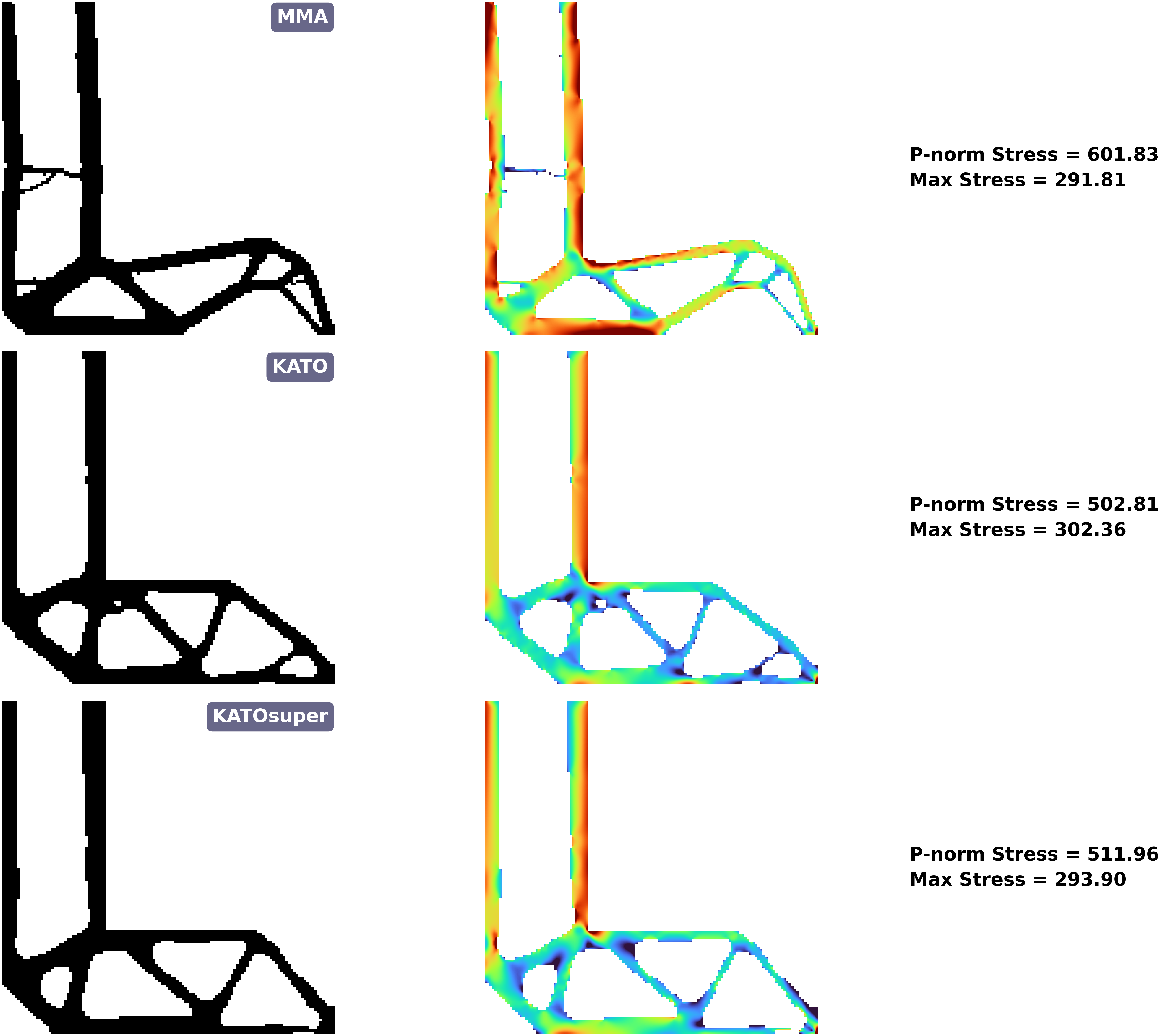}
  \caption{L-shaped bracket stress optimization results demonstrating stress concentration relief.}
  \label{fig:stress_lshape}
\end{figure}

Having established KATOsuper's effectiveness for compliance minimization, we now evaluate the more challenging stress optimization problem. Stress minimization presents a harder optimization landscape due to the highly nonlinear p-norm objective. We use realistic physical parameters (50 kN load, 10 mm thickness, steel material), reporting von Mises stress in MPa. Figs.~\ref{fig:stress_2d}--\ref{fig:stress_lshape} present the optimized structures with stress distributions.

The stress optimization results reveal KATO's advantages across benchmarks. For the cantilever beam (Fig.~\ref{fig:stress_2d}), KATOsuper achieves the best p-norm stress of 383.99 MPa and maximum von Mises stress of 115.58 MPa, outperforming both MMA (384.98/119.80 MPa) and KATO (384.26/117.90 MPa). On the MBB beam (Fig.~\ref{fig:stress_mbb}), results are more competitive: MMA achieves 398.89 MPa p-norm, KATO reaches 396.88 MPa (best p-norm), and KATOsuper obtains 397.55 MPa---all within 0.5\% of each other. For maximum von Mises stress on MBB beam, KATOsuper achieves 110.13 MPa, the lowest among all methods compared to MMA's 143.94 MPa. The most dramatic differences appear on the L-shaped bracket.

\rev{KATO methods significantly outperform MMA on the L-shaped bracket (Fig.~\ref{fig:stress_lshape}), reducing p-norm stress by 16.4\% (from 601.83 to 502.81 MPa). MMA achieves 601.83/291.80 MPa, KATO reaches 502.81/302.36 MPa (best p-norm), and KATOsuper obtains 511.96/293.90 MPa. The L-shape geometry presents the most challenging stress optimization scenario due to its re-entrant corner, which creates a geometric stress concentration. MMA's local update rule struggles to redistribute material globally to relieve this stress concentration, resulting in thin struts near the corner that amplify stress. In contrast, KATO's neural generator learns global material redistribution patterns that naturally avoid these high-stress configurations, producing thicker transitions and smoother load paths around the critical corner region.}

The relative performance differences between benchmarks reveal an important observation: KATO's advantage over conventional methods increases with problem complexity. On the relatively simple cantilever and MBB geometries, all methods achieve similar stress levels. On the geometrically complex L-shape bracket with stress concentration, KATO demonstrates substantial superiority. This suggests that neural reparameterization becomes increasingly valuable as optimization landscapes become more challenging with multiple local minima.

\rev{A subtle trade-off exists between the p-norm objective and maximum von Mises stress. This explains the apparent paradox in the L-bracket result, where KATOsuper attains lower maximum von Mises stress (293.90 MPa) than KATO (302.36 MPa) despite a higher p-norm stress (511.96 vs. 502.81 MPa). The p-norm is a global aggregate that weights all element stresses, not just the peak value. A design with slightly elevated stresses distributed across more elements can therefore exhibit a higher p-norm while simultaneously reducing the maximum stress at the critical location. In addition, the reported maximum von Mises stress is evaluated with singularity masking at load and boundary-condition locations, whereas the p-norm aggregates stresses over the full field, so the two metrics need not rank designs identically. Because KATO and KATOsuper converge to different local optima in a highly non-convex landscape, their trade-off between peak stress and p-norm aggregate differs.}

The stress sensitivity computation benefits substantially from automatic differentiation. Traditional MMA requires explicit derivation of the p-norm stress through nested loops over element stress components, resulting in complex implementations prone to numerical errors. In contrast, KATO computes $\partial \sigma_{\text{p-norm}} / \partial \boldsymbol{\rho}$ automatically through the computational graph, ensuring correctness and enabling efficient GPU parallelization. The simplified implementation facilitates the use of the same codebase for both compliance and stress objectives, requiring only a change in the loss function definition.

\begin{figure}[pos=h]
  \centering
  \includegraphics[width=0.6\textwidth]{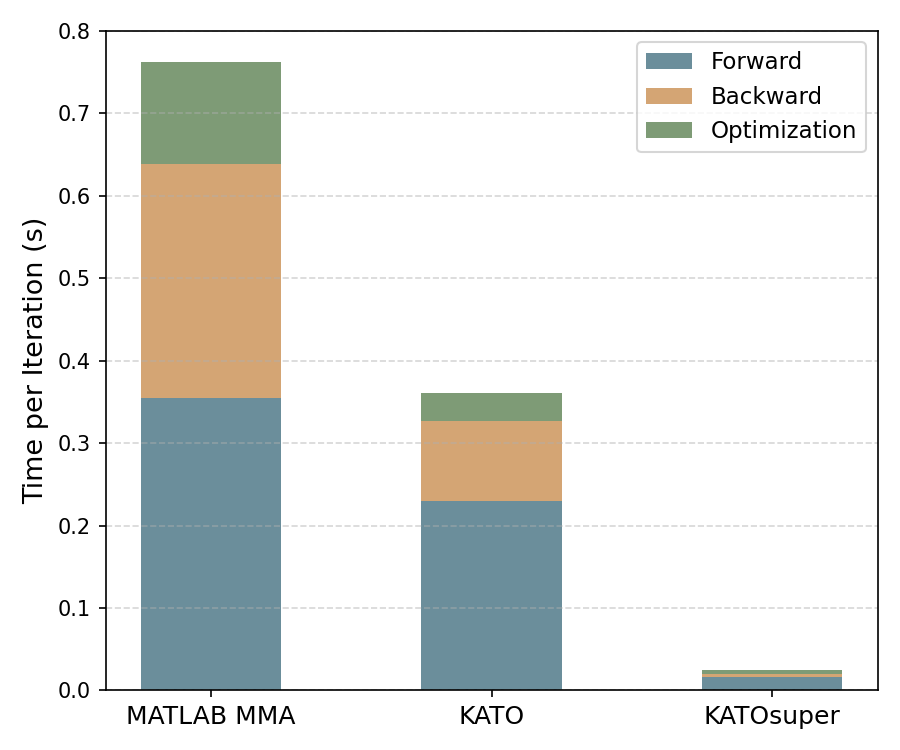}
  \caption{Computational time breakdown for 2D stress optimization (Cantilever with resolution of $128 \times 64$).}
  \label{fig:time_stress}
\end{figure}

The surrogate further accelerates stress optimization by eliminating the FEA solve entirely. Unlike compliance optimization where a single forward solve suffices, stress computation requires additional post-processing to compute strain and von Mises stress at each element. KATOsuper bypasses all of these operations, as illustrated in Fig.~\ref{fig:time_stress}, achieving \textbf{\revgreen{91$\times$} speedup} over MMA and \textbf{19$\times$ speedup} over KATO.

\subsection{Resolution extrapolation}
\begin{figure}[pos=h]
  \centering
  \includegraphics[width=0.6\textwidth]{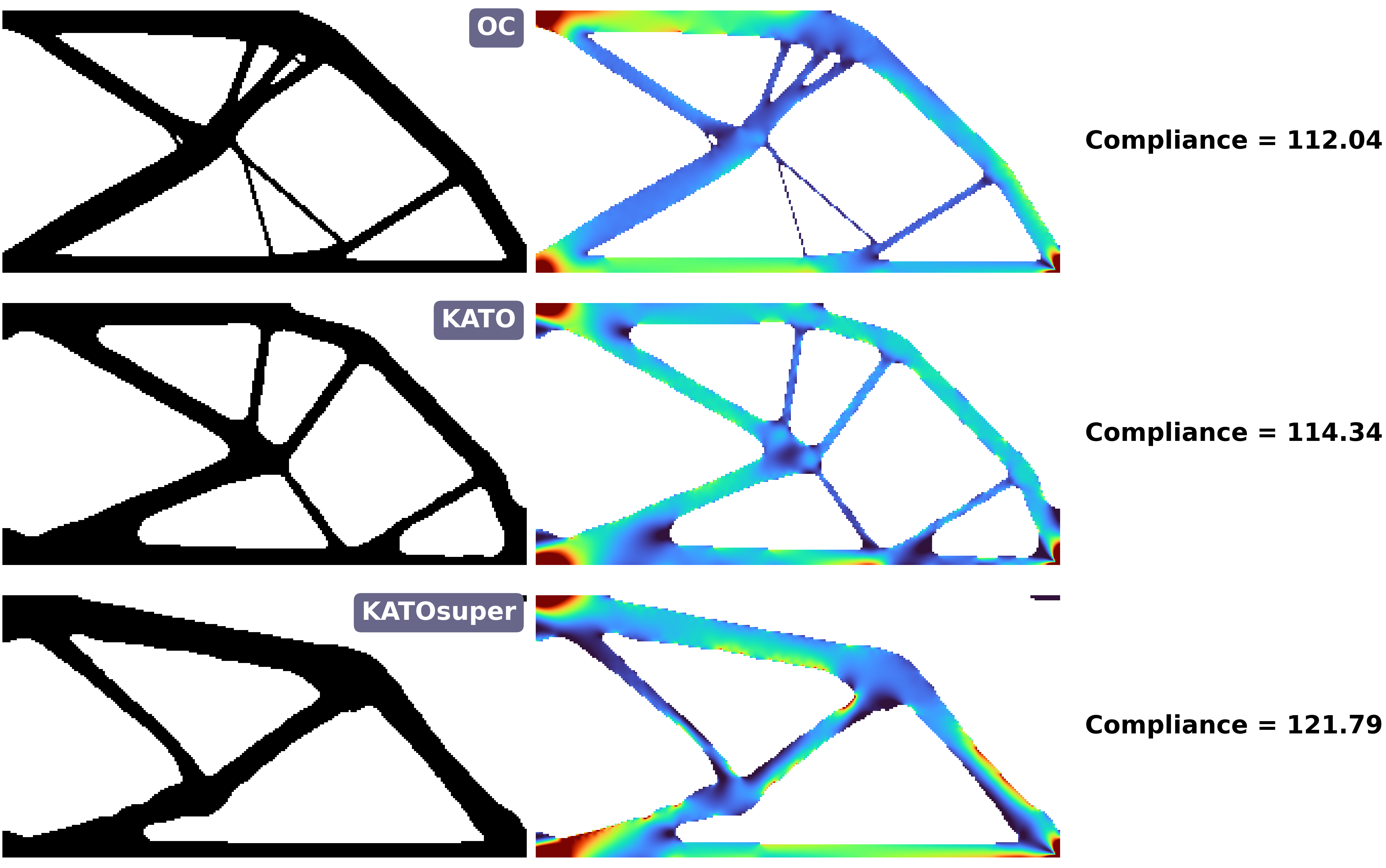}
  \caption{Cantilever resolution extrapolation at $256 \times 128$ (4$\times$ training resolution).}
  \label{fig:res_256}
\end{figure}

\begin{figure}[pos=h]
  \centering
  \includegraphics[width=0.6\textwidth]{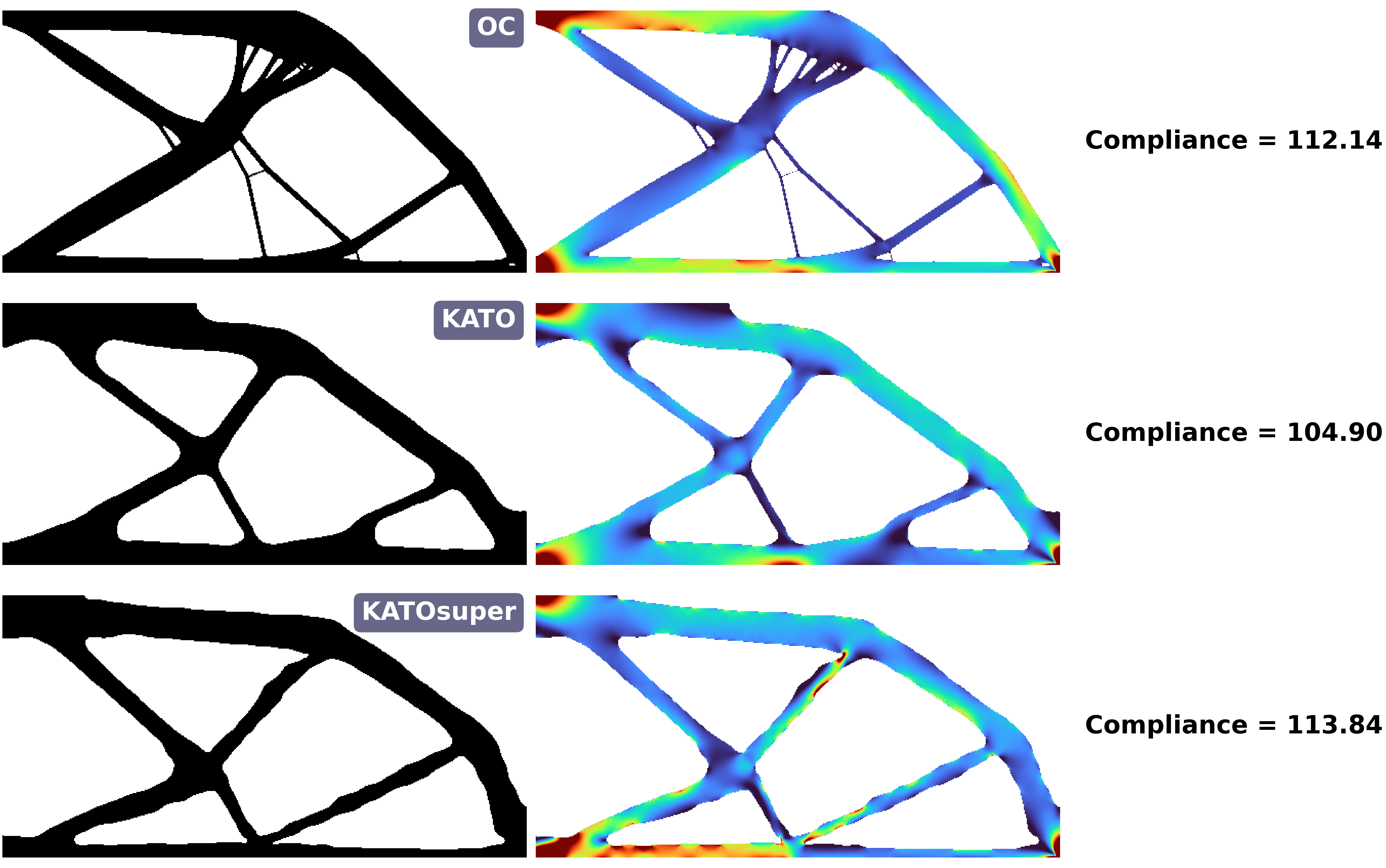}
  \caption{Cantilever resolution extrapolation at $512 \times 256$ (16$\times$ training resolution).}
  \label{fig:res_512}
\end{figure}

\begin{figure}[pos=h]
  \centering
  \includegraphics[width=0.6\textwidth]{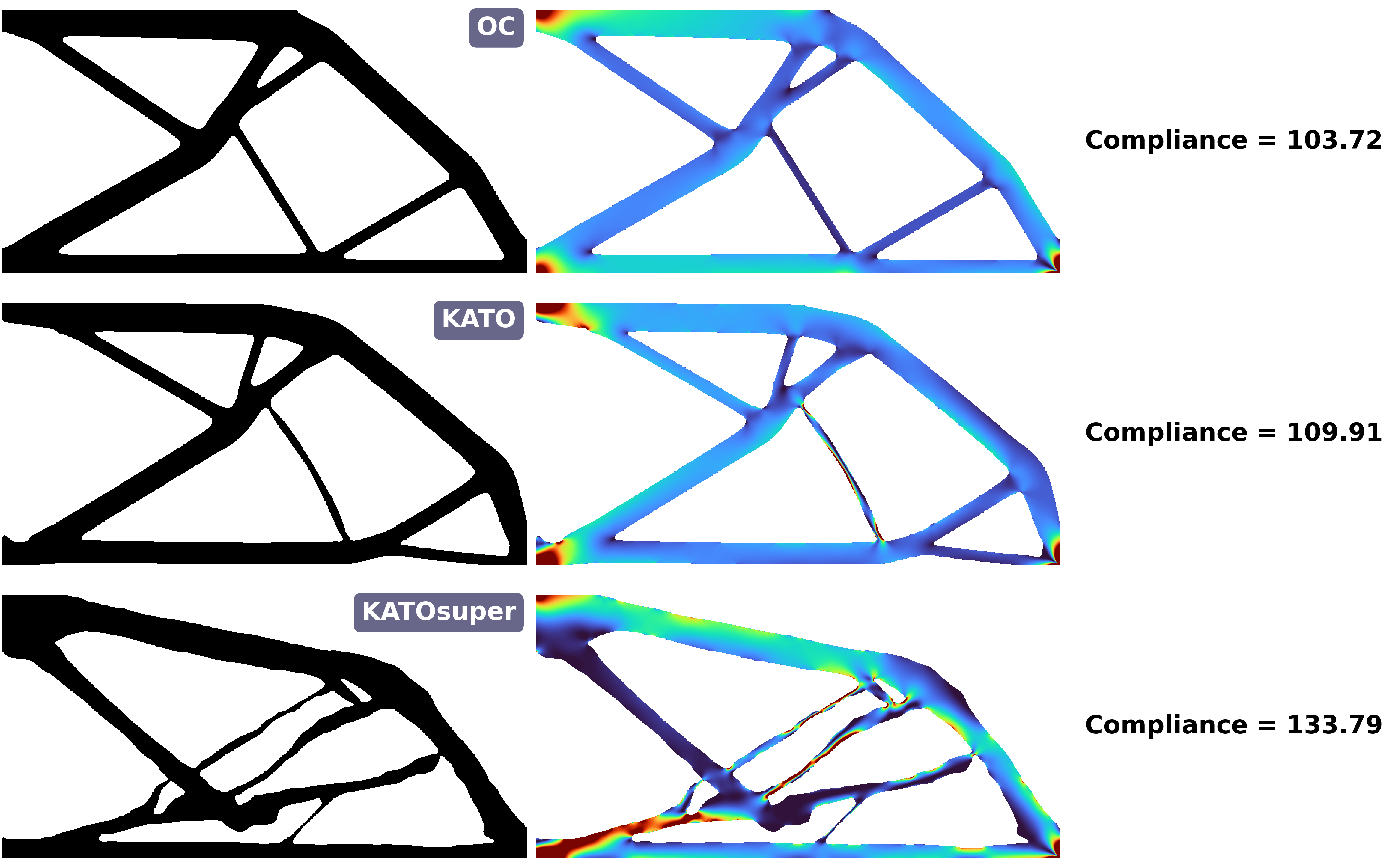}
  \caption{Cantilever resolution extrapolation at $1024 \times 512$ (64$\times$ training resolution).}
  \label{fig:res_1024}
\end{figure}

The preceding experiments used SC-FNO at its training resolution. A key advantage of the FNO architecture, however, is resolution-independent inference enabled by its spectral parameterization. We now evaluate zero-shot extrapolation by training SC-FNO at $128 \times 64$ and inferring at progressively higher resolutions. Figs.~\ref{fig:res_256}--\ref{fig:res_1024} present the optimized structures with compliance values.

At $4\times$ extrapolation (Fig.~\ref{fig:res_256}), OC achieves a compliance of 112.04, KATO obtains 114.34, and KATOsuper reaches 121.79. 
The 8.7\% gap between KATOsuper and OC highlights the difficulty of generalizing learned spectral representations to unseen resolutions without additional calibration.

At $16\times$ extrapolation (Fig.~\ref{fig:res_512}) with a resolution of $512 \times 256$, KATO achieves the best performance, outperforming OC by 6.5\% with a compliance of 104.90 compared to 112.14. 
KATOsuper yields a compliance of 113.84, remaining within \revgreen{8.5\%} of the KATO baseline despite the substantial resolution increase.
Notably, KATOsuper matches the classical OC method within a 1.5\% margin at this extrapolated resolution, indicating that the learned spectral representation preserves the essential low-frequency load-path structure under significant discretization shifts.

This behavior contrasts with conventional density-based approaches, where mesh refinement can induce pronounced topological variations or overly thin members unless carefully tuned filtering strategies are employed. 
By operating in Fourier space, SC-FNO emphasizes low-frequency modes that encode global structural behavior, which remain largely invariant across discretizations. 
As a result, the surrogate produces stable and topologically coherent designs when rendered at higher resolutions, even though fine-scale details are not explicitly learned during low-resolution training.

\revorange{While KATO gives the lowest compliance at $16\times$}, the optimization stability of KATOsuper under $16\times$ extrapolation is notable. 
Rather than replacing high-fidelity FEA, the surrogate provides directionally consistent sensitivity estimates that support stable optimization dynamics in high-resolution regimes, substantially reducing the computational cost associated with repeated large-scale FEA solves.
\revorange{The apparent ordering between the $4\times$ and $16\times$ examples should not be interpreted as a monotonic capability ranking. Each extrapolation level defines a separate nonconvex optimization trajectory, so isolated comparisons between two resolutions are less informative than the overall trend: extrapolation remains usable at moderate factors and degrades at high extrapolation, as the $64\times$ case in Fig.~\ref{fig:res_1024} shows.}

\rev{At 64$\times$ extrapolation ($1024 \times 512$), KATOsuper compliance increases to 133.79, representing 29\% degradation relative to OC (103.72). \revorange{This indicates that extrapolation quality worsens progressively with scaling factor.} However, the resulting structure remains topologically valid with clear load paths, indicating that the surrogate still provides useful directional guidance for design-space exploration. The structural topology at 64$\times$ extrapolation closely resembles the training-resolution optimum, with the primary difference being suboptimal fine-scale material distribution rather than fundamentally incorrect load-path identification.}

\rev{This usable-but-degraded performance enables a practical workflow: use KATOsuper for rapid design-space exploration at high resolution, then refine promising candidates with a few KATO iterations. The computational advantage remains significant (Table~\ref{tab:efficiency}): at 64$\times$ extrapolation, KATOsuper completes 50 iterations in 39s versus 2745s for MATLAB OC (70$\times$ speedup), making high-resolution exploration feasible even when final refinement requires FEA validation.}

\subsection{3D compliance optimization}
The resolution extrapolation experiments demonstrate KATOsuper's ability to generalize beyond training resolution in 2D. We now evaluate whether the framework extends to three-dimensional problems, where computational costs are substantially higher. We implement KATO3D, a pure KANConv3D generator operating on $64 \times 32 \times 4$ hexahedral meshes. The 3D extension presents unique challenges: training data generation becomes substantially more expensive (each 3D FEA solve requires approximately 10$\times$ the computation of 2D at equivalent mesh density), and the surrogate must capture more complex physics interactions including out-of-plane load redistribution.

\begin{figure}[pos=h]
  \centering
  \includegraphics[width=0.6\textwidth]{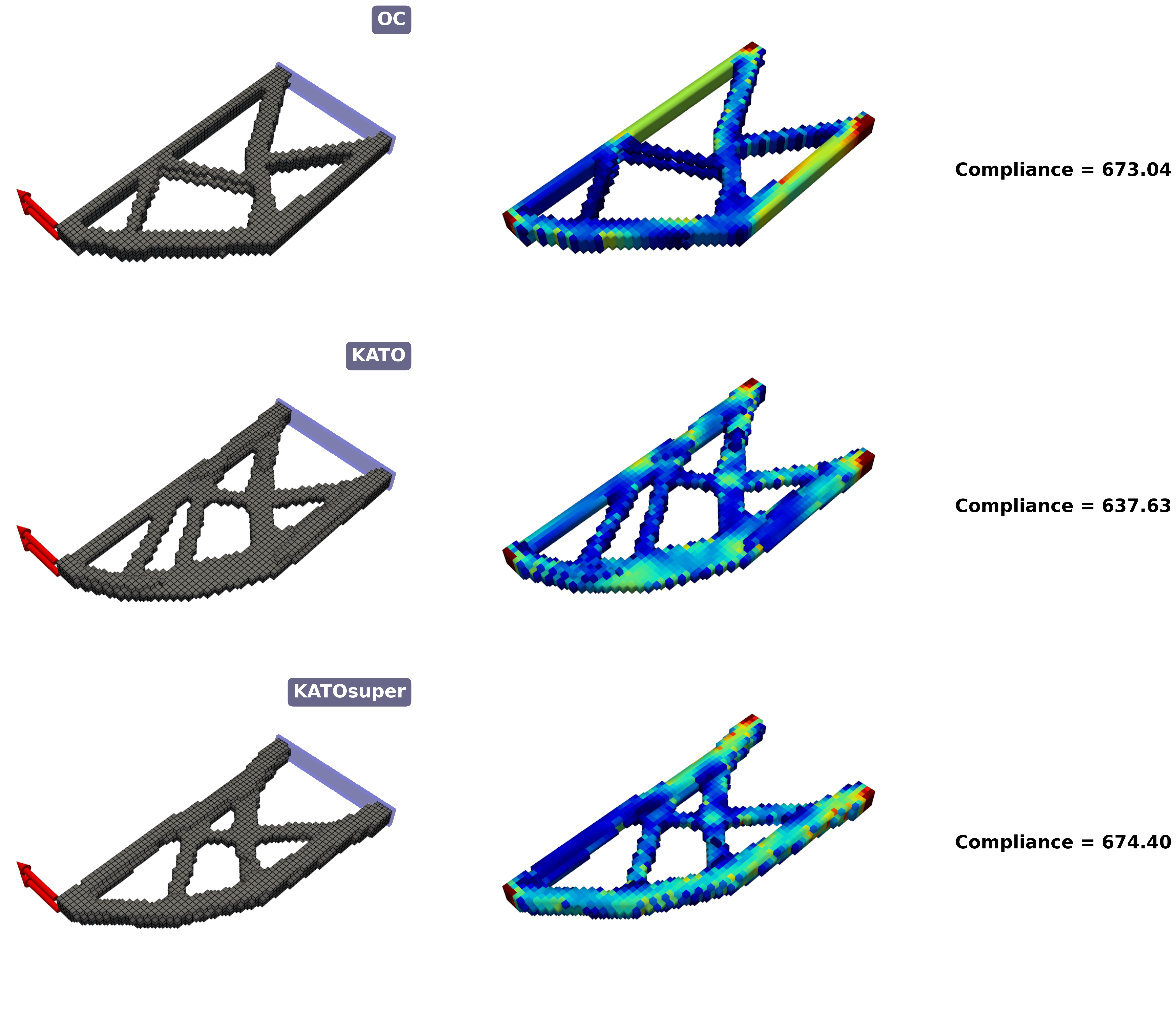}
  \caption{\revorange{Standard 3D cantilever ($64\times32\times4$). Rows: OC, KATO, and KATOsuper; leftmost column: boundary/load setup.}}
  \label{fig:3d_structure}
\end{figure}

\rev{Due to the reduced training data ($\sim$3,200 frames versus $\sim$6,200 for 2D)}, we employ online learning for KATOsuper to adapt the surrogate during optimization. Fig. \ref{fig:3d_structure} presents the optimized structures.

As shown in Fig.~\ref{fig:3d_structure}, KATO achieves compliance of 637.63, representing a 5.3\% improvement over OC's 673.04. KATOsuper with online learning achieves 674.40, closely matching OC. \revorange{These results indicate that the KANConv3D generator can capture the main 3D topology features in this benchmark.} The pure KAN architecture, with learnable B-spline activations, provides enhanced expressiveness for discovering complex 3D load paths. Unlike in 2D domains where multiple convolution types were tested, KATO3D uses exclusively KAN-based convolutions, \revorange{supporting their use in the tested volumetric setting}. The 3D structures exhibit proper load transfer through continuous plates and ribs, avoiding the disconnected material islands that sometimes appear in 2D optimizations.

\rev{The 3D experiments serve as a proof-of-concept demonstrating that the complete KATOsuper framework---neural reparameterization, surrogate evaluation, and online learning---functions correctly in the volumetric setting. The mesh sizes considered here are still modest relative to industrial-scale 3D topology optimization, because generating 3D training data remains substantially more expensive than in 2D.}

\begin{figure}[pos=h]
  \centering
  \includegraphics[width=0.6\textwidth]{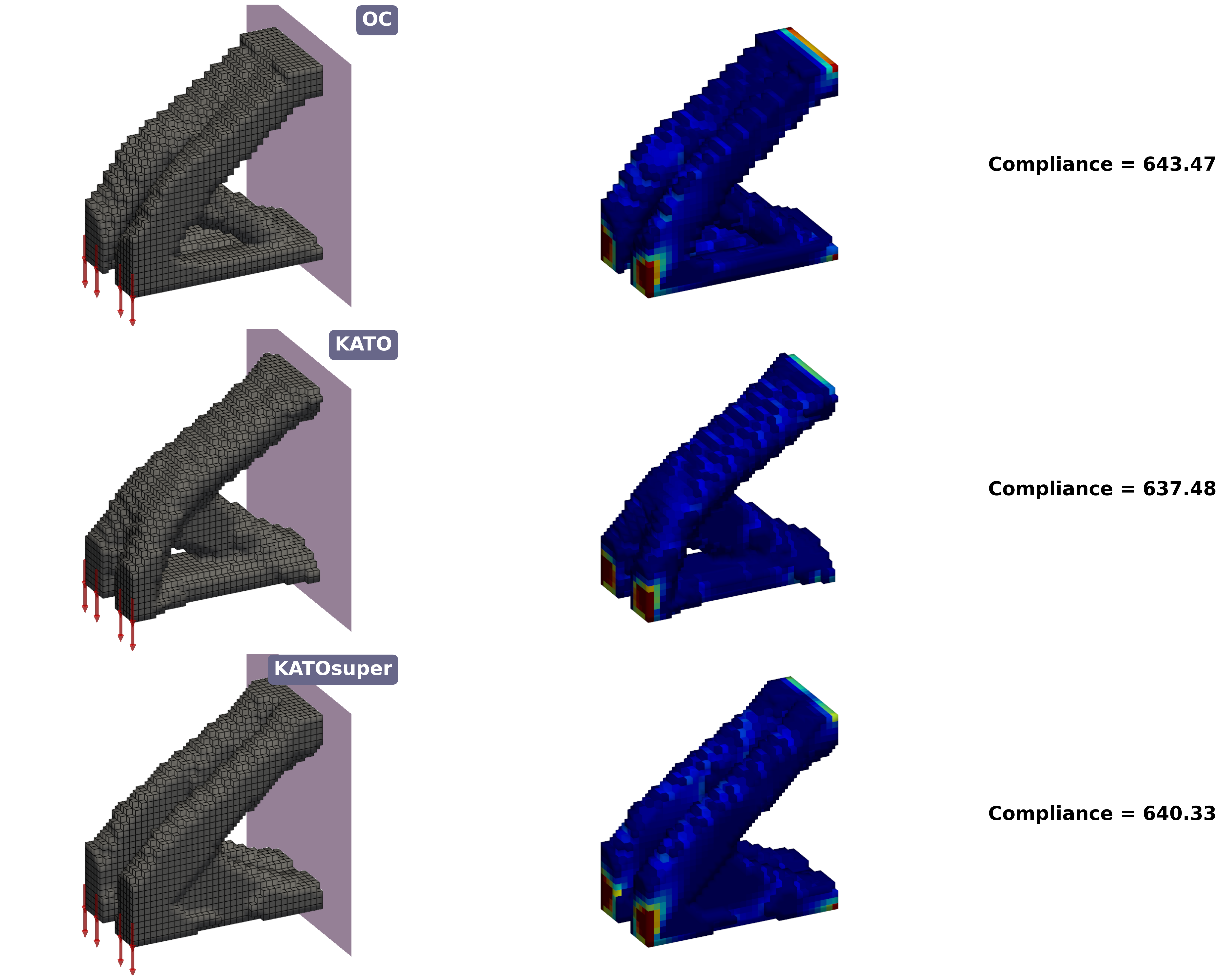}
  \caption{\revorange{3D split-load cantilever ($32\times32\times16$). Rows: OC, KATO, and KATOsuper; leftmost column: boundary/load setup.}}
  \label{fig:3d_split}
\end{figure}

\rev{To further validate KATO3D's capability on more complex geometries, we evaluate a \revorange{3D split-load cantilever} configuration at $32 \times 32 \times 16$ resolution, as shown in Fig.~\ref{fig:3d_split}. The downward load is split into two $5 \times 5$ surface patches located on the bottom faces at the front and back of the 16-element thickness domain, while fixed boundary conditions are applied to the right face. This asymmetric loading presents true volumetric optimization challenges with out-of-plane load redistribution.}

\begin{figure}[pos=h]
  \centering
  \includegraphics[width=0.9\textwidth]{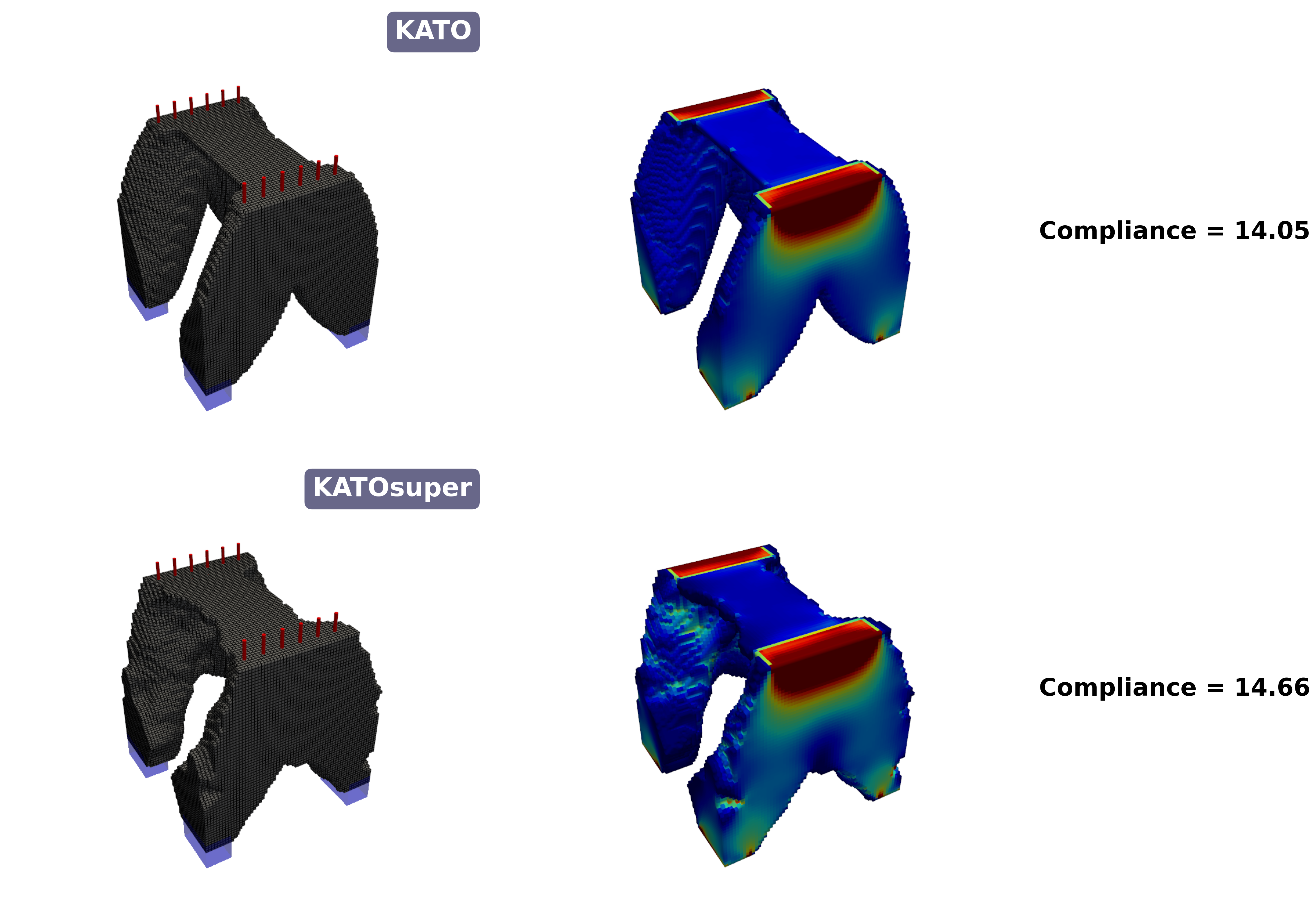}
  \caption{\revorange{Ship seat-base ($64\times32\times64$). Left/middle/right: structure, strain-energy density, and compliance; top/bottom rows: KATO and KATOsuper with online learning.}}
  \label{fig:shipseat_structure}
\end{figure}

For the \revorange{3D split-load cantilever} case (Fig.~\ref{fig:3d_split}), KATO gives compliance 637.48, compared with 643.47 for OC and 640.33 for KATOsuper. \revorange{The three values are close, with KATOsuper remaining in the same performance range as the FEA-based methods while providing substantial acceleration.} The angled structural members indicate that KATO3D captures 3D load paths that transfer forces through multiple planes.
\revorange{To evaluate larger-scale and new-geometry generalization, we also consider the ship seat-base ($64\times32\times64$, approximately 131k elements, volfrac 0.35) using the same 3D cantilever-trained surrogate with online FEA correction. The resulting topology and compliance values are shown in Fig.~\ref{fig:shipseat_structure}. This case is included as a supplementary check of online adaptation beyond the canonical cantilever/MBB/L-bracket geometries. In a 30-step run, KATOsuper uses 17 FEA calls rather than 30 for pure-FEA KATO. After binarization, largest-connected-component extraction, and FEA recomputation, KATOsuper reaches a compliance of 14.66, compared with 14.05 for KATO. The result indicates a connected and physically plausible ship seat-base structure with a modest compliance penalty relative to pure FEA.}

KATOsuper with online learning matches OC performance despite the reduced training data. The online learning strategy compensates for limited offline data by periodically injecting ground-truth FEA results during optimization. \revorange{The hybrid loop uses surrogate inference for most iterations and occasional FEA corrections to reduce error accumulation.} The injection frequency---comprising an initial 5-step burst followed by periodic calls every 4 iterations---balances correction frequency against computational overhead.

\begin{figure}[pos=h]
  \centering
  \includegraphics[width=0.6\textwidth]{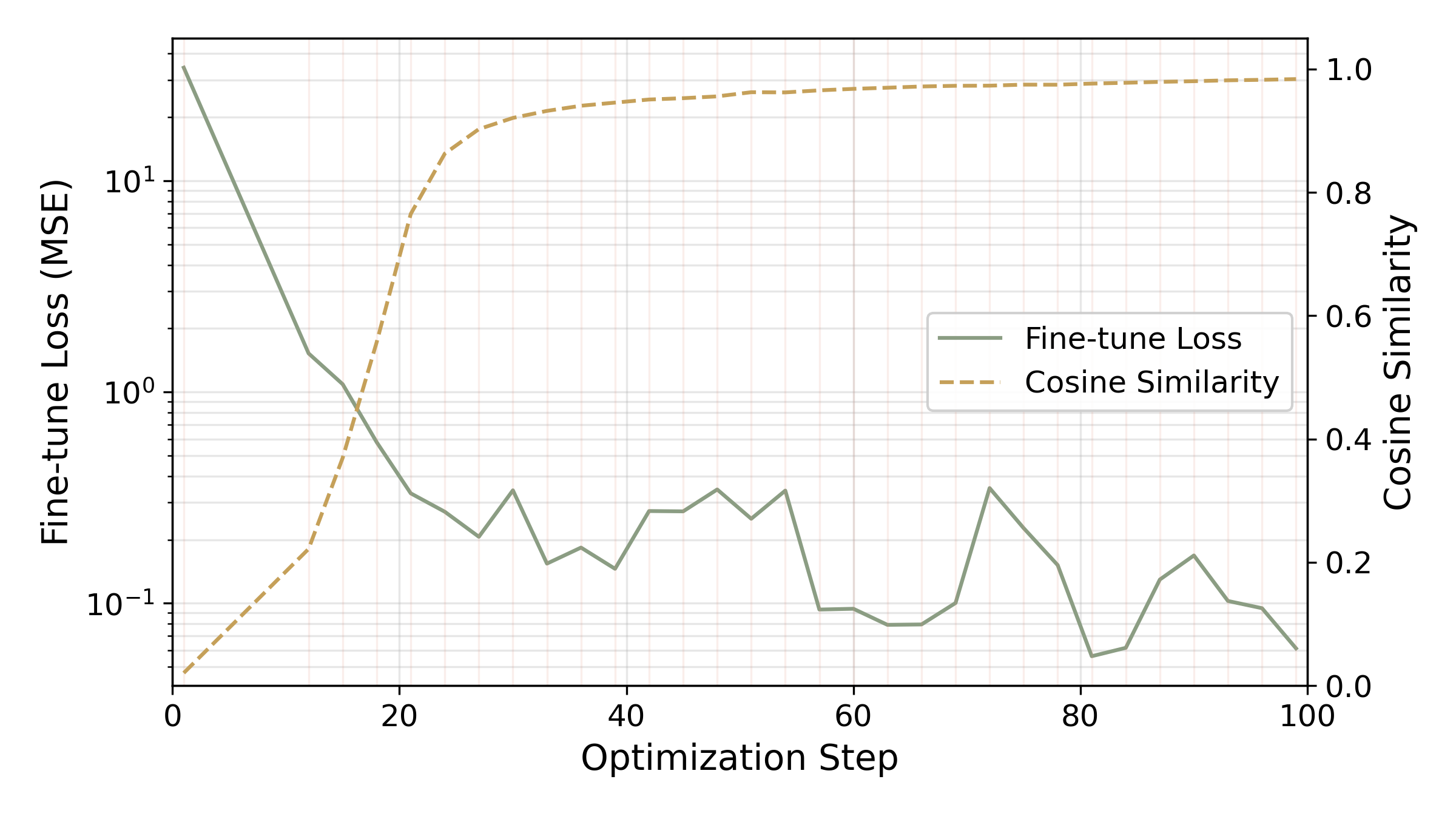}
  \caption{Online learning convergence during 3D cantilever ($64 \times 32 \times 4$) optimization: fine-tuning loss (left axis) and gradient cosine similarity (right axis) over optimization steps.}
  \label{fig:online_learning_curve}
\end{figure}

Fig.~\ref{fig:online_learning_curve} analyzes the convergence dynamics of the online surrogate during 3D optimization. The pre-trained SC-FNO model, trained on a limited 3D dataset, initially exhibits high prediction error (MSE $> 40$) when facing the random noise density fields generated by the KATO generator in the initial stages of the optimization. Consequently, the sensitivity direction derived via automatic differentiation lacks physical grounding, as evidenced by a gradient cosine similarity below 0.1. However, the online fine-tuning mechanism---executing calibration for the first five steps and every four steps thereafter---enables the surrogate to rapidly internalize the local physics manifold. By the total optimization step 24, corresponding to the 10th online learning intervention, the fine-tuning loss stabilizes below 0.2 while the sensitivity direction reaches near-perfect alignment with the ground truth ($\cos \theta \approx 0.95$). This rapid adaptation signifies that even a sparse pre-training baseline can be effectively leveraged to provide high-fidelity directional guidance once the design trajectory enters a physically plausible topology space.

\begin{figure}[pos=h]
  \centering
  \begin{subfigure}[t]{0.48\textwidth}
    \centering
    \includegraphics[width=\textwidth]{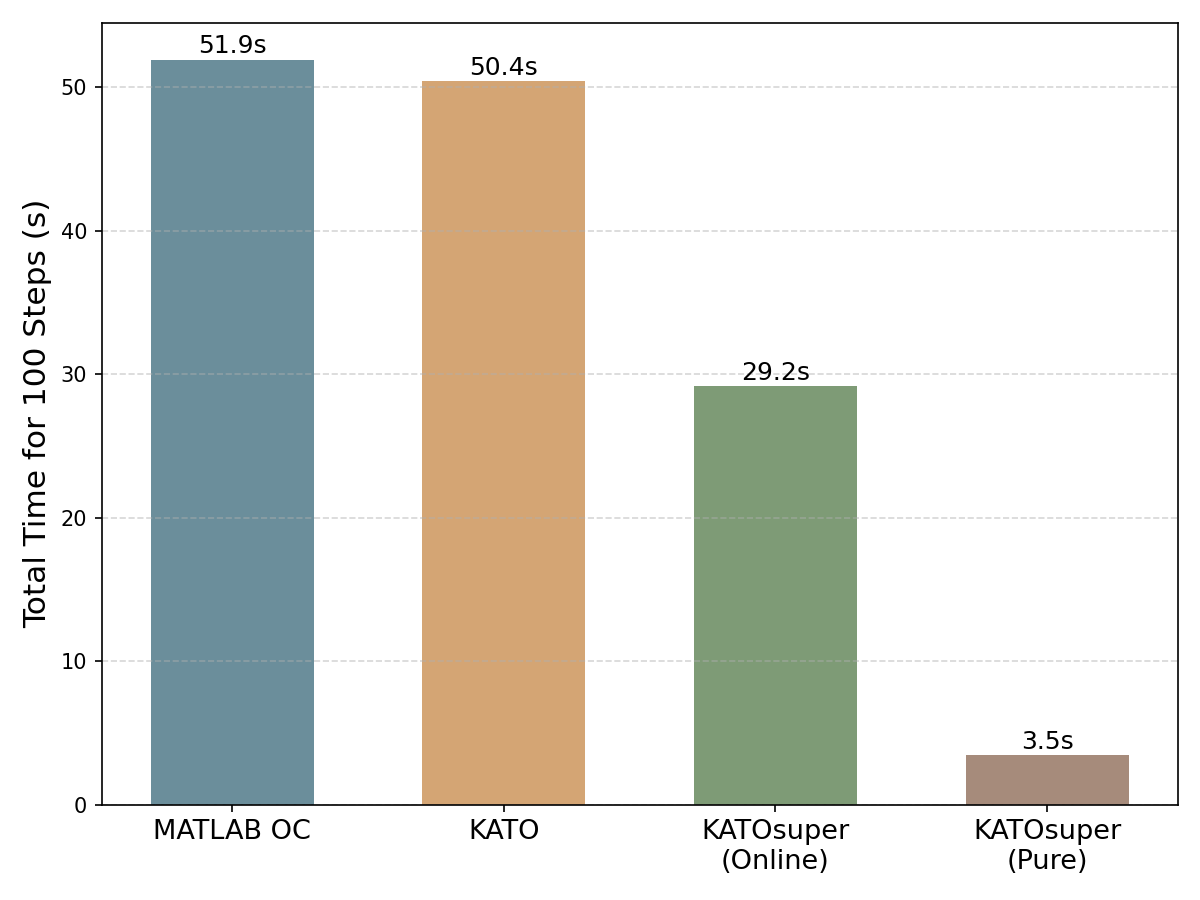}
    \caption{\revorange{Standard 3D cantilever ($64 \times 32 \times 4$)}}
  \end{subfigure}
  \hfill
  \begin{subfigure}[t]{0.48\textwidth}
    \centering
    \includegraphics[width=\textwidth]{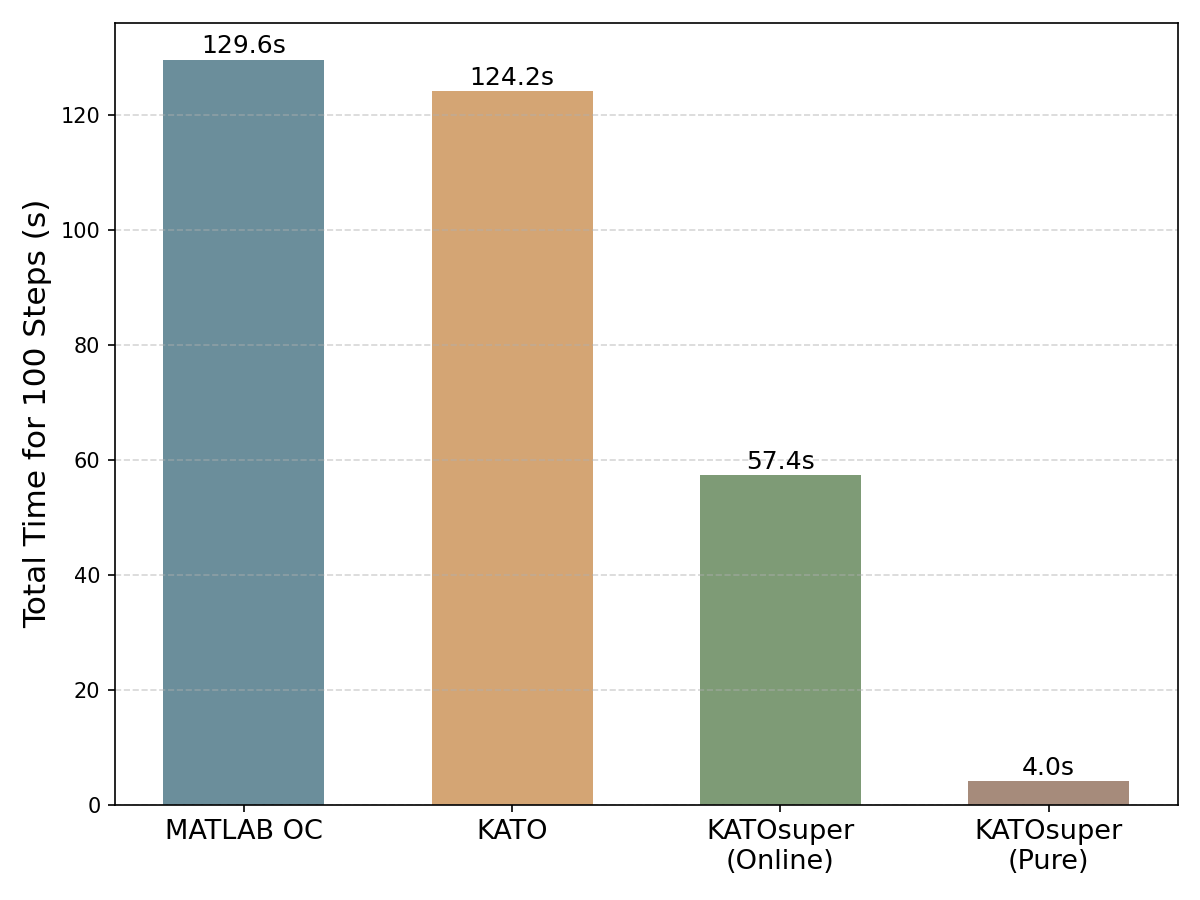}
    \caption{\revorange{3D split-load cantilever ($32 \times 32 \times 16$)}}
  \end{subfigure}
  \caption{Total optimization time comparison for 3D compliance problems (50 iterations).}
  \label{fig:online_convergence}
\end{figure}

The efficiency gains of KATOsuper become increasingly pronounced as the degrees of freedom (DOF) increase, as illustrated in Fig.~\ref{fig:online_convergence}. In \revorange{this moderate 3D online-learning setting}, the standard cantilever case ($64 \times 32 \times 4$) \revorange{uses periodic FEA correction} and halves the total optimization time compared to the MATLAB baseline (29.2s vs. 51.9s). When the element count doubles in the \revorange{3D split-load cantilever} configuration ($32 \times 32 \times 16$), the absolute wall-time advantage expands further, reducing design time from 129.6s to 57.4s (a 2.2$\times$ speedup). \revorange{These online-hybrid timings are separate from the deployment timings in Table~\ref{tab:efficiency} and from the additional generalization tests where more frequent FEA correction is used for different boundary conditions or geometries.} This trend reflects the fundamental scaling disparity between FEA and neural operators: as discretization refines, the $O(N^{1.5})$ complexity of sparse solvers leads to a super-linear cost increase, whereas the $O(N \log N)$ spectral convolutions of SC-FNO maintain high throughput. \revorange{The results show that online calibration can recover local directional accuracy while still reducing the number of expensive 3D factorizations.}

\subsection{Efficiency summary}
Table~\ref{tab:efficiency} summarizes computational performance across all experiments, comparing MATLAB baselines, \revgreen{an open-source Python baseline,} KATO (\revorange{neural-reparameterized with FEA calls}), and KATOsuper (surrogate-accelerated KATO). \revorange{The Python column uses the same OC/MMA formulations with SciPy's open-source sparse linear algebra on the same CPU.} \revgreen{KATO uses a hybrid configuration in which neural generator/autograd operations run on GPU and FEA runs on CPU; KATOsuper inference runs on the RTX 4070 GPU. The reported values are per-optimization deployment times, while the online-hybrid 3D timing with periodic FEA correction is analyzed separately in Fig.~\ref{fig:online_convergence}.}

\begin{table*}[t]
\centering
\caption{\revorange{Deployment-time computational efficiency comparison across benchmark cases.}}
\label{tab:efficiency}
\small
\resizebox{\textwidth}{!}{%
\begin{tabular}{@{}llcccccc@{}}
\hline
Task & Resolution & \makecell{\revorange{OC/MMA}\\\revorange{with MATLAB}} & \makecell{\revorange{OC/MMA}\\\revorange{with Python}} & KATO & KATOsuper & \makecell{\revorange{KATOsuper}\\\revorange{speedup vs.}\\\revorange{OC/MMA with MATLAB}} & \makecell{\revorange{KATOsuper}\\\revorange{speedup vs.}\\\revorange{KATO}} \\
\hline
2D Compliance & $128 \times 64$ & 14.2 s & \revgreen{79.4 s} & 16.7 s & 0.75 s & \textbf{19$\times$} & \textbf{22$\times$} \\
2D Compliance & $256 \times 128$ & 99 s & \revgreen{1450 s} & 67 s & 1.3 s & \textbf{76$\times$} & \textbf{52$\times$} \\
2D Compliance & $512 \times 256$ & 413 s & \revgreen{10260 s} & 269 s & 3.75 s & \textbf{110$\times$} & \textbf{72$\times$} \\
2D Compliance & $1024 \times 512$ & 2745 s & \revgreen{OOM} & 4010 s* & 39 s & \textbf{70$\times$} & \textbf{103$\times$} \\
2D Stress & $128 \times 64$ & \revgreen{76.6 s} & \revgreen{318 s} & 16.3 s & 0.84 s & \textbf{\revgreen{91$\times$}} & \textbf{19$\times$} \\
3D Compliance & $64 \times 32 \times 4$ & 51.9 s & \revgreen{81 s} & 50.4 s & 3.5 s & \textbf{15$\times$} & \textbf{14$\times$} \\
3D Compliance & $32 \times 32 \times 16$ & 129.6 s & \revgreen{873 s} & 124.2 s & 4.0 s & \textbf{32$\times$} & \textbf{31$\times$} \\
\hline
\multicolumn{8}{l}{\footnotesize \rev{*KATO 1024$\times$512 time inflated due to PARDISO memory overflow.} \revorange{OOM denotes out-of-memory failure.}}
\end{tabular}}
\end{table*}

\rev{The surrogate-accelerated framework achieves \revorange{15--110$\times$ deployment-time speedup over MATLAB OC/MMA baselines}, with greater gains at higher resolutions due to the FNO's $O(N \log N)$ complexity versus FEA's $O(N^{1.5})$ sparse solver scaling. At $512 \times 256$ resolution, KATOsuper achieves 110$\times$ speedup over the FEA baseline. \revorange{In the deployment timing table, the 3D case exhibits the same pattern: KATOsuper runs 15$\times$ faster than MATLAB OC while KATO only matches MATLAB speed, highlighting the practical value of surrogate acceleration for 3D applications where FEA cost dominates.}} \revgreen{The Python baseline utilizes the open-source implementations from Refs.~\cite{liu2014efficient,andreassen2011efficient}. This Python version exhibits lower computational efficiency compared to the MATLAB baseline and suffers from out-of-memory (OOM) errors at a resolution of $1024 \times 512$. These results demonstrate that the MATLAB implementation serves as a competitive and rigorous benchmark rather than an artificially weakened baseline.} \revgreen{KATOsuper remains faster than both MATLAB and the open-source same-language Python baseline in all completed cases.}

\begin{table*}[t]
\centering
\caption{\revorange{Peak CPU RAM and GPU VRAM (GB) across timing benchmarks.}}
\label{tab:memory}
\scriptsize
\begin{tabular*}{\textwidth}{@{\extracolsep{\fill}}llcccc@{}}
\hline
Task & Resolution & MATLAB CPU & Python CPU & KATO CPU/GPU & KATOsuper CPU/GPU \\
\hline
2D Compliance & $128 \times 64$ & 1.00 & 0.59 & 1.40/0.19 & 1.69/0.61 \\
2D Compliance & $256 \times 128$ & 2.04 & 3.84 & 1.74/0.68 & 1.95/0.94 \\
2D Compliance & $512 \times 256$ & 6.72 & 19.04 & 2.73/2.63 & 2.92/2.89 \\
2D Compliance & $1024 \times 512$ & 24.27 & OOM & 7.48/10.37 & 7.34/10.72 \\
2D Stress & $128 \times 64$ & 1.57 & 0.68 & 1.37/0.22 & 1.64/0.63 \\
3D Compliance & $64 \times 32 \times 4$ & 1.29 & 0.52 & 1.80/0.75 & 2.36/2.25 \\
3D Compliance & $32 \times 32 \times 16$ & 3.92 & 1.76 & 2.77/1.47 & 3.34/2.97 \\
\hline
\multicolumn{6}{l}{\footnotesize \revorange{OOM denotes out-of-memory failure.}}\\
\end{tabular*}
\end{table*}
\revorange{Table~\ref{tab:memory} reports peak CPU RAM and GPU VRAM for the same cases.} \revgreen{The results show the same pattern as the timing comparison. The direct FEA baselines are limited by sparse-factorization storage: MATLAB reaches 24.27\,GB at $1024\times512$, while the open-source Python baseline fails on the same workstation. KATO and KATOsuper move the neural components to GPU memory and keep CPU memory below 8\,GB in the largest 2D case, with KATOsuper using 10.72\,GB of GPU memory. Thus, the acceleration in Table~\ref{tab:efficiency} is accompanied by a shift from CPU sparse-solver memory pressure to bounded GPU tensor memory, which is the intended deployment regime for the surrogate evaluator.}

\begin{figure}[pos=h]
  \centering
  \includegraphics[width=0.6\textwidth]{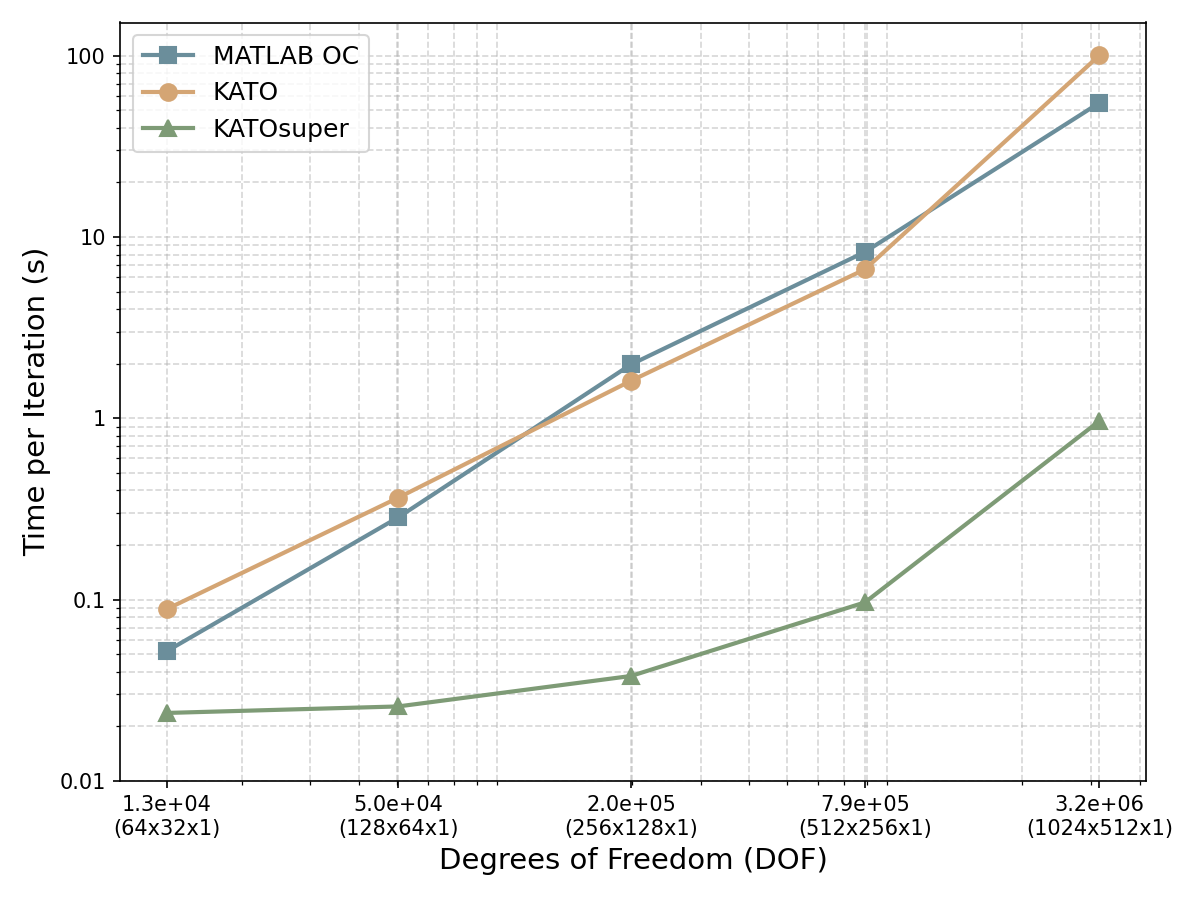}
  \caption{Computational complexity comparison (Cantilever) between KATOsuper and FEA-based methods across resolutions.}
  \label{fig:scalability}
\end{figure}

Fig.~\ref{fig:scalability} visualizes the end-to-end time scaling of the optimization loop. While KATOsuper exhibits orders-of-magnitude lower absolute runtime, its curve bends upward at the highest resolution ($1024 \times 512$). This behavior stems from how GPU workloads transition from being dominated by fixed overheads to being dominated by memory traffic.

At low-to-moderate resolutions the GPU is under-utilized, so wall time is dominated by relatively resolution-insensitive overheads such as kernel launch and framework dispatch. This makes the curve appear nearly flat from $64$ to $512$ elements. When the resolution reaches $1024 \times 512$, the workload becomes large enough that data movement and arithmetic dominate, producing a visually sharp knee in the scaling curve. In contrast, CPU-based KATO/FEA starts from a much higher baseline and therefore appears to grow more smoothly on a log-log scale.

At extreme resolution, KATOsuper's bottleneck shifts from the surrogate evaluator to the generator. The default generator ends with a KAN head that evaluates B-spline basis functions per pixel, producing large intermediate tensors whose size scales with the number of pixels. At $1024 \times 512$ resolution (approximately $5.2 \times 10^5$ pixels), activation expansion inflates tensors to hundreds of megabytes temporary. During backpropagation these intermediates must be stored and read, pushing the computation into a memory-bound regime where cache efficiency drops and bandwidth becomes the limiting factor. This explains the nonlinear GPU runtime increase at the highest resolution even though the method remains far from compute-bound. Compared with the previous KATO implementation that required an 80GB A100 GPU at $512 \times 256$ resolution \cite{yan2025kato}, the current PyTorch implementation runs stably at $1024 \times 512$ using \revgreen{10.72\,GB} of GPU memory in the reported measurement.

This analysis reveals a key difference in where computation is spent. MATLAB OC and FEA-based KATO are dominated by the sparse linear solver, whose practical scaling is super-linear (often $O(N^{1.5})$ to $O(N^2)$ depending on fill-in and memory). KATOsuper removes this solver bottleneck but reveals generator-side memory traffic, particularly from KAN layers, as the next limiting factor. This is a deliberate trade-off: KAN-based generation improves expressiveness and yields smoother, more physically plausible topology boundaries, while still providing more than two orders of magnitude speed advantage \revred{compared with the FEA-based KATO baseline} at $1024 \times 512$ resolution.

\section{Discussion}\label{sec:discussion}

\subsection{Sensitivity consistency as the key enabler}
\textit{Sensitivity direction} matters more than magnitude accuracy, particularly when dealing with complex objective functions. In TO, the complexity of the sensitivity field varies significantly with the choice of objective. Compliance minimization is a relatively smooth global task where the sensitivity field is broadly distributed; in such cases, dual-head architectures that independently regress compliance and sensitivity can sometimes produce acceptable optimization trajectories despite inherent mathematical inconsistencies.

However, the necessity of SC-FNO's consistent architecture becomes evident when optimizing highly non-linear and non-convex objectives like p-norm stress. Stress sensitivity fields are characterized by extreme localization, high-frequency signals, and singular points---locations at geometric discontinuities (e.g., re-entrant corners or load application points) where stress concentrations approach theoretical infinity. Direct regression of such fields is prone to ``gradient pollution''---the predicted sensitivities exhibit sign errors and local noise that are mathematically decoupled from the predicted stress values. Our early experimental observations confirmed that independent-head designs fail catastrophically in stress optimization, often diverging within a few iterations or producing meaningless topologies. This failure highlights that sensitivity consistency is not merely a performance enhancement but a fundamental requirement for stable design synthesis under complex physical objectives.

\rev{SC-FNO's \texttt{forward\_split} mechanism solves this by reforming the problem: the network learns to predict the lower-complexity scalar field (stress or compliance) directly, and the intricate sensitivity peaks are recovered through automatic differentiation. This approach ensures that the deployed sensitivity follows the same first-order computational path as the predicted objective field, providing the optimizer with a physically coherent descent direction even when the physics evaluation is approximate. By shifting from field regression to field differentiation, KATOsuper provides the neural generator with a more reliable signal for load-path discovery across diverse physical design tasks.}

\subsection{Computational efficiency and scalability}
\rev{KATOsuper achieves \revorange{15--110$\times$ deployment-time speedup over MATLAB OC/MMA baselines}, with speedup increasing at higher resolutions. This favorable scaling behavior reflects a fundamental shift in computational paradigm: from solving linear systems to evaluating learned operators.}

\rev{Traditional sparse direct solvers such as PARDISO exhibit memory complexity of $O(N^{\beta})$ with $\beta \approx 1.5$--$2$, arising from fill-in during matrix factorization. The factorization step dominates computational cost and scales super-linearly with \revred{problem size}. In contrast, SC-FNO operates in spectral space via FFT, achieving $O(N \log N)$ computational complexity and $O(N)$ memory footprint. At \revred{$1024 \times 512$ resolution}, this difference translates to 103$\times$ acceleration \revred{compared with the FEA-based KATO baseline}. We note that KATOsuper and GPU-accelerated FEA frameworks such as JAX-FEM \cite{xue2023jax} address computational efficiency at different levels. JAX-FEM accelerates the FEA solver itself via GPU parallelism and automatic differentiation, whereas KATOsuper replaces repeated FEA evaluations with a learned operator. These approaches are not mutually exclusive: SC-FNO training data could be generated using JAX-FEM or other accelerated solvers, which would reduce the offline data-generation burden without changing the basic deployment principle of replacing repeated FEA calls with operator inference.}

The log-log scalability plot (Fig.~\ref{fig:scalability}) reveals this gap: FEA methods exhibit steeper slopes indicative of super-linear scaling, while KATOsuper maintains near-linear growth. Beyond the training resolution, SC-FNO's complexity advantage becomes increasingly pronounced, enabling optimization at resolutions impractical for conventional methods.

\subsubsection{The physical memory wall}
At \revred{$1024 \times 512$ resolution}, FEA solve time exhibits a dramatic non-linear jump. When increasing from $512 \times 256$ to $1024 \times 512$, PARDISO iteration time increases from approximately 3 seconds to over 87 seconds---a 29$\times$ increase for merely 4$\times$ more elements. This phenomenon reflects the memory wall: when sparse matrix storage exceeds available cache, page swapping introduces severe latency. Direct solvers such as SciPy's LU factorization encounter this limit at even lower resolutions due to dense fill-in patterns.

\revred{SC-FNO avoids this sparse-factorization barrier.} Neural operators store only network parameters (approximately 2M weights for our architecture), independent of mesh resolution. \revred{The forward pass scales through FFT-based tensor operations rather than sparse matrix construction and factorization, avoiding the solver storage bottleneck that dominates direct FEA.} This architectural difference establishes surrogate-based optimization as a viable path toward full-scale 3D topology optimization, where memory constraints otherwise preclude practical computation.

\subsubsection{End-to-end GPU acceleration}
The efficiency advantage extends beyond the physics solver. In KATO, the neural generator runs on GPU while FEA executes on CPU, incurring data transfer latency at each iteration. \revgreen{A CPU/GPU generator microbenchmark gives forward--backward times of 22.6/5.5\,ms, 88.2/9.3\,ms, and 443.8/39.6\,ms at $128\times64$, $256\times128$, and $512\times256$, respectively, while the density-transfer cost remains below 0.15\,ms. If the generator is forced to run on CPU, generator inference alone requires approximately 3.5 seconds at $1024 \times 512$ resolution; in the reported KATO configuration, this generator component is run on GPU while FEA remains on CPU.} KATOsuper eliminates this bottleneck by maintaining the entire pipeline---generator and SC-FNO---on GPU. Tensor parallelism across thousands of CUDA cores reduces generator inference to milliseconds, maintaining high throughput even at extreme resolutions.

\subsubsection{PyTorch implementation advantages}
Relative to our prior work KATO(legacy) \cite{yan2025kato}, the PyTorch-based reimplementation and architectural modifications in KATOsuper achieve substantial efficiency gains. KATO(legacy) was implemented in TensorFlow and tested on NVIDIA A100 (80GB vRAM), whereas the current KATOsuper runs on NVIDIA RTX 4070 (12GB vRAM)---a GPU with approximately 25\% of the computational throughput. Despite this hardware disparity, per-iteration time at $128 \times 64$ resolution decreased from 0.36s to 0.19s, representing a 1.88$\times$ speedup. More significantly, the maximum stable resolution increased from approximately 40,000 elements in KATO(legacy) to over 520,000 elements ($1024 \times 512$) in KATOsuper---a 13$\times$ improvement in scalability.

This efficiency gain stems from three factors: (1) PyTorch's optimized Autograd implementation with superior memory management; (2) vectorized tensor operations using \texttt{einsum} that eliminate explicit loops; and (3) reduced vRAM consumption through careful intermediate tensor handling. The result is a framework that achieves higher performance on consumer-grade hardware, dramatically lowering the entry barrier for practitioners without access to data center GPUs.

\subsubsection{Automatic differentiation for sensitivity analysis}
Stress optimization requires sensitivity of the p-norm aggregated von Mises stress with respect to element densities. Deriving manual adjoint equations for this highly nonlinear objective is algebraically complex and could suffer from inappropriate implementation. SC-FNO computes $\partial \sigma_{\text{p-norm}} / \partial \boldsymbol{\rho}$ automatically through backpropagation, enabling gradient-based optimization for arbitrary differentiable objectives without case-specific derivation. This flexibility extends naturally to multi-physics objectives including thermal compliance or coupled thermo-mechanical problems.

\subsubsection{Online learning for practical 3D applications}
\rev{For 3D problems where offline training data generation becomes prohibitively expensive, online learning provides a balanced solution. Pure surrogate inference may accumulate prediction errors over many iterations, while pure FEA remains too slow for interactive design. The hybrid strategy---surrogate inference with periodic FEA correction every 4 iterations---achieves both physical accuracy and computational efficiency. As shown in Fig.~\ref{fig:online_convergence}, total optimization time reduces by about 50\% relative to pure FEA, while gradient cosine similarity remains high throughout optimization. The oscillating fine-tune loss observed after step 40 is consistent with partial forgetting across successive calibrations: each calibration step adapts the surrogate to the current density configuration and may partially overwrite knowledge from previous configurations. Despite this oscillation in loss magnitude, the gradient direction remains robustly aligned. Within roughly half of the pure-FEA runtime, the surrogate model is fine-tuned to better represent the specific scenario and resolution; in our 3D compliance setting, this adaptation also transfers across related runs, such as moving from the \revorange{standard 3D cantilever configuration} at $64 \times 32 \times 4$ to the \revorange{3D split-load cantilever} at $32 \times 32 \times 16$.} \revorange{In a separate online on/off check on the 3D split-load cantilever, the uncalibrated surrogate drifts to a poor, FEA-unverified design, whereas online calibration raises the gradient cosine similarity from 0.07 to 0.99 and returns the design to the FEA-consistent performance range. For the reported 3D split-load cantilever in Fig.~\ref{fig:3d_split}, the formal benchmark values are KATO 637.48, OC 643.47, and KATOsuper 640.33. The adapted surrogate can then be reused in new optimization runs without further FEA correction and remains stable when returned to the standard 3D cantilever configuration. This supports the interpretation of online learning as adaptation to new boundary conditions rather than zero-shot boundary-condition extrapolation.}

Table~\ref{tab:paradigm} summarizes the paradigm shift from traditional FEA to neural operator-based optimization.

\begin{table*}[t]
\centering
\caption{\rev{Comparison of solver paradigms for topology optimization. This table summarizes the configurations considered in this work and is not intended as an exhaustive survey of all possible GPU-accelerated or AD-enabled FEA implementations.}}
\label{tab:paradigm}
\small
\resizebox{\textwidth}{!}{%
\begin{tabular}{@{}lp{2.4cm}p{2.5cm}p{2.5cm}p{6.0cm}@{}}
\hline
\textbf{Aspect} & \textbf{Vanilla FEA} & \textbf{PARDISO} & \textbf{SC-FNO} & \textbf{Insight} \\
\hline
Memory & $O(N^{2-3})$ & $O(N^{1.5-2})$ & $O(N)$ & Avoids factorization storage \\
Computation & $O(N^{2.5})$ & $O(N^{1.5})$ & $O(N \log N)$ & Near-linear via FFT \\
Sensitivity & Manual adjoint & Manual adjoint & Auto-diff & Arbitrary objectives \\
Hardware & CPU only & CPU multi-core & GPU parallel & Modern accelerator fit \\
\hline
\end{tabular}}
\end{table*}

\rev{The training cost deserves explicit analysis. Offline preparation involves two stages: data generation (running KATO optimization trajectories to produce training frames) and SC-FNO network training. For the 2D compliance surrogate, 154 KATO runs at $128 \times 64$ resolution produce $\sim$6,200 training frames in approximately 0.5\,h of CPU time; SC-FNO training then requires $\sim$4.8\,h on a single GPU. For the 3D surrogate, 80 runs at $64 \times 32 \times 4$ ($\sim$50\,s each) produce $\sim$3,200 frames in $\sim$1.1\,h, with training requiring $\sim$1.9\,h. Complete offline costs are summarized in Table~\ref{tab:config}. Data generation is embarrassingly parallel and can be further accelerated by distributing runs across multiple CPU cores.}
\revgreen{Using the open-source Python baseline in Table~\ref{tab:efficiency}, the one-time 2D offline cost ($\sim$19,080\,s) is recovered after approximately 243 optimizations at $128\times64$, 13 at $256\times128$, and 2 at $512\times256$; at $1024\times512$, the Python baseline runs out of memory on the same workstation.}

\rev{In practical design studies, large numbers of FEA solves are routinely performed during exploration, parametric sweeps, and iterative optimization workflows. These computations naturally generate physics--design pairs across varying resolutions, boundary conditions, and load configurations. Owing to the resolution-agnostic formulation of SC-FNO and its ability to perform zero-shot resolution extrapolation, such heterogeneous FEA outputs can be accumulated and reused as training data without requiring a carefully curated resolution-specific dataset.}

\rev{Online learning further reduces reliance on large offline datasets by periodically adapting the surrogate during optimization. In this mode, periodic FEA calls recalibrate the surrogate on the encountered trajectory, mitigating distribution shift and improving local fidelity for related out-of-distribution configurations.}
\revgreen{We do not add a 2D $64\times$ online-learning case because 2D data generation is comparatively cheap and the trained 2D surrogate already supports zero-shot high-resolution exploration; online calibration is mainly needed in 3D, where offline data generation is much more expensive.}

\subsection{Limitations}
Several limitations warrant discussion. First, zero-shot resolution extrapolation exhibits an inherent performance-generalization trade-off. While FNO is theoretically discretization-invariant, we observe a 29\% increase in compliance at 64$\times$ extrapolation compared to ground-truth OC results. This accuracy degradation is not a failure of the network architecture but can be explained through the lens of information theory~\cite{shannon1949communication}. 

Following the Nyquist-Shannon sampling theorem~\cite{shannon1949communication}, the training resolution ($128 \times 64$) defines the maximum physical mode frequencies the neural operator can capture. When inferring at $1024 \times 512$, the surrogate essentially acts as a spectral low-pass filter; it accurately captures the low-frequency ``structural skeleton'' and primary load paths, but cannot represent the high-frequency physical details that emerge at high discretization levels. This spectral aliasing leads to the observed gap in numerical optimality, though the resulting structures remain topologically valid and connectivity-preserving.

Second, the SC-FNO requires data generation for offline training, which involves an upfront computational cost (running KATO optimization trajectories). This amortizes well when many optimizations are performed, but for domains where only a few optimizations are needed, the pure KATO approach (without surrogate) may remain preferable.

Third, the surrogate assumes the test problem lies within the training distribution. Boundary condition configurations, load patterns, or material properties substantially different from training may produce poor predictions. The online learning mechanism partially addresses this by injecting FEA corrections, but at the cost of reduced speedup. Explicit uncertainty quantification would enable automatic detection of out-of-distribution cases.

Fourth, the current online learning implementation uses a fixed injection schedule (every 4 iterations). Adaptive scheduling based on prediction confidence could improve efficiency, injecting FEA only when needed. This remains an avenue for future improvement.

\subsection{Future directions}
Several extensions merit investigation. Multi-physics objectives (e.g., thermo-mechanical coupling) could leverage the same surrogate framework by expanding the input encoding to include additional field quantities such as temperature distributions or thermal conductivities. The SC-FNO architecture remains applicable---the key requirement is that the objective can be computed from a forward pass through the physics solver, enabling automatic differentiation.

Uncertainty quantification through ensemble methods~\cite{lakshminarayanan2017simple} or Bayesian neural operators~\cite{psaros2023uncertainty} would enable automatic detection of out-of-distribution inputs, triggering FEA fallback only when necessary. This would make KATOsuper more robust in production environments where users may specify configurations not covered by training.
\revgreen{The uncertainty clustering, NSF/sensitivity filtering, and load-location surrogate modeling explored in recent robust level-set and concurrent topology--layout studies~\cite{li2026cmame_multiscale_uncertainty,li2026ast_device_layout} suggest a natural extension of KATOsuper toward uncertainty-aware surrogate calibration and load/configuration selection.}

The decoupled generator-evaluator architecture naturally supports transfer learning~\cite{pan2009survey}: a generator trained on one domain could be fine-tuned for related problems, while the surrogate could be retrained independently. For example, a generator trained on cantilever beams might transfer to bridge/ship deck optimization with minimal fine-tuning, while the surrogate learns bridge-specific physics. This modularity may accelerate deployment across diverse engineering applications including additive manufacturing design~\cite{jihong2021review}, compliant mechanism synthesis, and manufacturing-constrained design \cite{vatanabe2016topology}.

\section{Conclusion}\label{sec:conclusion}
\rev{This work presented KATOsuper, a unified framework coupling neural-reparameterized topology optimization with Sensitivity-Consistent Fourier Neural Operators. By addressing both the computational bottleneck of repeated FEA solves and the gradient consistency requirements of surrogate-driven optimization, KATOsuper enables practical acceleration of topology optimization across multiple dimensions and objectives. The key contributions are:}

\begin{itemize}
    \item \rev{\textbf{KATO3D}: \revorange{A 3D extension of neural-reparameterized topology optimization using a latent-space KANConv3D generator. In the tested 3D cantilever case, KATO gives 5.3\% lower compliance than OC (C=638 vs C=674), while the overall 3D results indicate competitive optimization quality.} The present 3D experiments serve as a proof-of-concept for volumetric surrogate-accelerated topology optimization.}
    
    \item \rev{\textbf{SC-FNO evaluator design}: A Sensitivity-Consistent Fourier Neural Operator combining the sensitivity-constrained training principle of \cite{behroozi2025sensitivity} with a TO-specific \texttt{forward\_split} inference architecture. Unlike dual-head surrogates that regress objective and sensitivity independently, \texttt{forward\_split} separates density inputs (gradient-tracked) from context channels (detached), ensuring that the deployed sensitivity is obtained as $\partial J_{\text{pred}} / \partial \boldsymbol{\rho}$ via automatic differentiation through the predicted objective field. The same evaluator design can be instantiated for different differentiable objectives.}
    
    \item \rev{\textbf{Resolution extrapolation}: FNO's spectral design enables zero-shot inference beyond the training resolution, up to $64\times$ in our experiments (train on $128 \times 64$, optimize at $1024 \times 512$) without retraining. Performance degrades progressively beyond the training resolution due to the fundamental spectral bandwidth limitation, reaching a 29\% compliance increase at $64\times$. Nevertheless, the extrapolated structures remain topologically valid and suitable for design-space exploration prior to refinement.}
    
    \item \rev{\textbf{Online learning}: A training and adaptation strategy in which periodic FEA calls during optimization recalibrate the surrogate model. The cosine similarity loss term preserves gradient direction alignment, which proves more important than exact magnitude accuracy for successful optimization.}
    
    \item \rev{\textbf{Multi-objective capability}: The same framework design supports compliance and stress minimization in 2D, and extends to 3D compliance through KATO3D. For stress optimization, KATO significantly outperforms MMA on geometrically complex cases (L-shape: 502 vs 602 MPa p-norm stress, 16.4\% reduction).}
    
    \item \rev{\textbf{Computational efficiency}: \revorange{15--110$\times$ deployment-time speedup over MATLAB OC/MMA baselines}, with greater gains at higher resolutions where FEA memory scaling dominates. At $512 \times 256$ resolution, KATOsuper achieves 110$\times$ acceleration.}
    
    \item \rev{\textbf{Small-data efficiency}: Strong performance with modest training data ($\sim$6,200 frames for 2D, $\sim$3,200 frames for 3D), demonstrating that online learning can compensate for limited offline data in the tested settings.}
\end{itemize}

KATOsuper \revgreen{maintains competitive optimality in the tested settings} while dramatically reducing computational cost. The core insight---that sensitivity \textit{direction} matters more than magnitude---enables robust optimization even with approximate physics evaluation. This principle may generalize beyond topology optimization to other physics-driven design problems where exact gradients are expensive but directional guidance suffices.

\rev{KATOsuper provides a scalable framework for surrogate-accelerated neural topology optimization. The decoupled generator-evaluator architecture supports modular improvements: generators can be fine-tuned for specific applications while surrogates adapt to new physics domains. As the framework matures, we anticipate extensions to multi-physics objectives, uncertainty-aware optimization, and integration with manufacturing constraints~\cite{vatanabe2016topology} for end-to-end design automation.}

%\section*{\rev{Data and Code Availability}}
%\rev{The complete KATOsuper implementation, including the generator, SC-FNO surrogate, training scripts, and \revgreen{main configuration files for the reported benchmarks}, is publicly available at \url{https://github.com/ysyysy115/KATOsuper}.}

\appendix

\section*{Acknowledgement}
The authors acknowledge the financial support by Natural Sciences and Engineering Research Council of Canada (NSERC) [grant number RGPIN-2025-04421 and DGDND-2025-04421]. This research was supported in part by computational resources and services provided by Advanced Research Computing at The University of British Columbia. The authors also gratefully acknowledge Google Research for providing fundamental methodological inspiration~\cite{hoyer2019neural} that enabled the development of the neural topology optimization framework.

%\section*{Declaration of generative AI and AI-assisted technologies in the writing process}
%During the preparation of this work, the authors used ChatGPT (OpenAI) in order to enhance readability and assist with code development. After using this tool, the authors reviewed and edited the content as needed and take full responsibility for the content of the publication.

\bibliographystyle{unsrtnat}
\bibliography{cas-refs}

\end{document}